\documentclass[traditabstract]{aa}
\usepackage{graphicx}
\usepackage{amsmath}
\usepackage[varg]{txfonts}
\usepackage[sort]{natbib}
\usepackage{color}
\bibpunct{(}{)}{;}{a}{}{,}
\newcommand\anchor[2]{#2}%
\newcommand\url{\@dblarg\@url}%
\def\@url[#1]{\anchor{#1}}%
\def\@text@email#1{#1}%
\newcommand\phn{\phantom{0}}%
\newcommand\phd{\phantom{.}}%
\newcommand{\myemail}{CSandin@aip.de}
\def\iraf{\textsc{iraf}}
\def\p3d{\textsc{p3d}}

\def\rH{H86}
\def\rS{S06}
\def\rSB{S10}

\begin{document}
\title{\p3d: a general data-reduction tool for fiber-fed\\ integral-field spectrographs\thanks{Based in part on observations collected at the Centro Astron\'omico Hispano Alem\'an (CAHA), operated jointly by the Max-Planck Institut f\"ur Astronomie and the Instituto de Astrof\'isica de Andalucia (CSIC).}}

\titlerunning{\p3d: a general data-reduction tool for fiber-fed IFUs}

\author{C.\ Sandin\inst{1}
  \and T.\ Becker\inst{1}
  \and M.~M.\ Roth\inst{1}
  \and J.\ Gerssen\inst{1}
  \and A.\ Monreal-Ibero\inst{2}
  \and P.\ B\"ohm\inst{1}
  \and P.\ Weilbacher\inst{1}}
\institute{Astrophysikalisches Institut Potsdam (AIP), An der Sternwarte 16, D-14482 Potsdam, Germany
  \and European Organisation for Astronomical Research in the Southern Hemisphere (ESO), Karl-Schwarzschild-Stra{\ss}e 2, D-85748 Garching bei M{\"u}nchen, Germany}

\offprints{\myemail}

\date{Received January 8, 2010 / Accepted February 11, 2010}

\abstract{The reduction of integral-field spectrograph (IFS) data is demanding work. Many repetitive operations are required in order to convert raw data into, typically a large number of, spectra. This effort can be markedly simplified through the use of a tool or pipeline, which is designed to complete many of the repetitive operations without human interaction. Here we present our semi-automatic data-reduction tool {\p3d} that is designed to be used with fiber-fed IFSs. Important components of {\p3d} include a novel algorithm for automatic finding and tracing of spectra on the detector, and two methods of optimal spectrum extraction in addition to standard aperture extraction. {\p3d} also provides tools to combine several images, perform wavelength calibration and flat field data. {\p3d} is at the moment configured for four IFSs. In order to evaluate its performance we have tested the different components of the tool. For these tests we used both simulated and observational data. We demonstrate that for three of the IFSs a correction for so-called cross-talk due to overlapping spectra on the detector is required. Without such a correction spectra will be inaccurate, in particular if there is a significant intensity gradient across the object. Our tests showed that {\p3d} is able to produce accurate results. {\p3d} is a highly general and freely available tool. It is easily extended to include improved algorithms, new visualization tools and support for additional instruments. The program code can be downloaded from the {\p3d}-project web site \anchor{http://p3d.sourceforge.net}{http://p3d.sourceforge.net}.}


\maketitle

\section{Introduction}\label{sec:introduction}
With integral integral field spectrographs (IFSs) an extended area on the sky can be spectroscopically mapped, under the same observing conditions, in one single exposure. In order to fit all simultaneously observed spectra onto the detector, the field-of-view of the integral field unit (IFU) that provides the spatial sampling on the sky is relatively small. The footprint of IFUs typically ranges from several arcsec, e.g.\ PMAS, to about one arcmin for VIMOS. In an alternative configuration, IFUs are used to map a much larger area, but with sparse sampling, e.g.\ PPAK and VIRUS-P. In Table~\ref{sandint1} we list several of the existing fiber-fed IFUs, the telescope they are mounted on, and the names of the corresponding pipelines. See \citet{Be:09} for a complete list. Regardless of the actual IFS configuration, the raw data of fiber-fed IFSs always consist of hundreds, or even thousands, of spectra per exposure. The reduction of these data consists of processing each spectrum individually and is therefore highly repetitive work. A general data-reduction tool is highly desirable, that automates the reduction steps, yet allows the user to interactively inspect and optimize parameters when required. The purpose of {\p3d} is to provide such capabilities and thereby facilitate the scientific exploitation of IFSs.

\begin{table*}[t]
\caption{A list of fiber-fed IFUs and their respective data-reduction pipelines}
\label{sandint1}
\tabcolsep=10pt
\begin{tabular}{ll@{\hspace{10pt}}l@{\hspace{10pt}}cclc}
\hline\hline
\noalign{\smallskip}
Telescope & \multicolumn{1}{c}{Spectrograph} & IFU & $n_\text{d}$ & Ref. & Reduction tool/Pipeline & Ref.\\[1.5pt]
\hline
\noalign{\smallskip}
VLT/UT2 & GIRAFFE & FLAMES-ARGUS   & 1 & 1 & \textsc{bldrs} & 1a\\
        &         &                &   &   & \textsc{giraffe pipeline} & 1b\\
Gemini North/South && GMOS-N, GMOS-S &1& 2 &                &   \\
Magellan I         && IMACS          &8& 3 &                & 3a\\
WHT                &WYFFOS& INTEGRAL       &1& 4 \\
Calar Alto 3.5m    &PMAS &LARR      &1& 5 & \textsc{P3d}, \textsc{p3d\_online}  & 5a\\
                   &     &PPAK      &1& 6 & \textsc{ppak\_online}\\
AAT & AAOMEGA      & SPIRAL         &2& 7 & \textsc{2dfdr} & 7a\\
VLT/UT3 &VIMOS     & VIMOS-IFU      &4& 8 & \textsc{vipgi} & 8a\\
      &&             &&& \textsc{vimos pipeline}\\
McDonald 2.7m      &VIRUS-P& VIRUS-P     &1   &  9 & \textsc{vaccine} &9a\\[1.0ex]
\hline
\noalign{\smallskip}
\end{tabular}\\
\textbf{Notes.}\ The name of the telescope, the spectrograph and the IFU are given in Cols.~1--3. In Col.~4 we specify the number of detectors of the IFU. Column~5 specifies the main instrument reference paper, and Cols.~6--7 give the name and reference of instrument-specific reduction tools/pipelines.\\[1.0ex]
\textbf{References.}\ $^1$\citet{AvGuJo.:03}, $^{1\text{a}}$\citet{BlCaNo.:00}, $^{1\text{b}}$\citet{PaAvAl.:00}; $^2$\citet{AlMuCo.:02}; $^3$\citet{ScDoCo.:04}, $^{3\text{a}}$\citet{BoBu:07}; $^4$\citet{ArCaCa.:98}; $^5$\citet{RoKeFe.:05}, $^{5\text{a}}$\citet{TBe:02}; $^6$\citet{KeVeRo.:06}; $^7$\citet{SmSaBr.:04}, $^{7\text{a}}$\citet{ShSaSm.:06}; $^{8}$\citet{LeSaMa.:03}, $^{8\text{a}}$\citet{ScFrGa.:05}; $^{9}$\citet{HiMaSm.:08}, $^{9\text{a}}$\citet{Adetal:10}.
\end{table*}

Special purpose data-reduction pipelines exist for most IFUs. Two additional reduction packages that are more general in their functionality than the ones listed in Table~\ref{sandint1}, and are suitable to use with fiber-fed IFUs, are \textsc{r3d} \citep[hereafter \rS]{Sa:06} and {\iraf}\footnote{The Image Reduction and Analysis Facility {\iraf} is distributed by the National Optical Astronomy Observatories which is operated by the association of Universities for Research in Astronomy, Inc.\ under cooperative agreement with the National Science Foundation.}. These tools, however, require significant amounts of time-consuming manual interaction in order to get the best out of them. While developed initially for the PMAS spectrograph, the data-reduction tasks of {\p3d} (and \textsc{p3d\_online}) are very generally formulated and are equally applicable to data obtained with other fiber-fed IFSs. The suitability of {\p3d} for several other IFSs is demonstrated with a variety of scientific results at an early stage, when no such tools were available yet for those instruments shortly after commissioning, e.g.\ MPFS (\citealt{RoBeKe.:04}; \citealt{LeBeFa.:05}; \citealt{FaShBe.:05}) or VIMOS \citep{MoRoSc.:05,ViSaDe.:06}.

In this paper we present a generalized version of {\p3d}, that is a complete rewrite of the previous version of \citet{TBe:02}. This general IFS reduction tool includes support for PMAS/LARR, PMAS/PPAK, VIRUS-P and SPIRAL, and can be readily extended to additional IFUs. Key features of {\p3d} are:
\begin{itemize}
\item A single, freely available, and easy to install, program package with support for several IFSs and computing platforms.
\item Calculation and propagation of errors through all steps.
\item The option to choose between aperture extraction and two methods of optimal extraction. 
\item A possibility to store information about all performed operations in a log file.
\item Interactive and integrated visualization tools to allow the user to examine intermediate and final products.
\end{itemize}
The code, furthermore, includes full run-time error handling and both the code and supplementary data files are fully commented.

This paper is laid out as follows. In Sect.~\ref{sec:program} we first describe the goals and setup of {\p3d}. The data-reduction algorithms and their implementation are thereafter detailed in Sect.~\ref{sec:datareduction}. We discuss and analyze the outcome of the tool, and make a comparison with corresponding outcome of {\iraf}, for some of the tasks, in Sect.~\ref{sec:dataanalysis}. Finally, we close the paper with our conclusions in Sect.~\ref{sec:conclusions}.

\section{About the objective and setup of {\p3d}}\label{sec:program}
Our goal with {\p3d} is to provide a general, flexible, fast and reliable tool for reduction of fiber-fed IFU data. {\p3d} is not intended to be a tool that handles every possible task from data reduction to data analysis. Instead it focuses on the tedious and repetitive tasks, which are required to convert raw data into wavelength-calibrated spectra, cf.\ Sect.~\ref{sec:datareduction}. We also want to provide a user-friendly tool where the required input from the user is kept to a minimum. It is recommended, however, that the user has a basic knowledge of the data-reduction process, and the related numerical problems and methods, in order to allow a full exploitation of all benefits of {\p3d}.

The collection of routines that make up {\p3d} were from the start \citep{TBe:02,RoKeFe.:05} written in the Interactive Data Language\footnote{\anchor{http://www.ittvis.com}{http://www.ittvis.com}} (IDL). Supplementary routines, which are mainly related to file-IO, are used from the publicly available toolkit astro-lib of NASA\footnote{\anchor{http://idlastro.gsfc.nasa.gov}{http://idlastro.gsfc.nasa.gov}}. The \textsc{mpfit} routine of \citet{Ma:09}\footnote{\anchor{http://purl.com/net/mpfit}{http://purl.com/net/mpfit}} is also used when fitting line profiles (for the optimal spectrum extraction). The current version of {\p3d} -- that is a public release under GPL-v3 -- is a complete rewrite of earlier versions and remains a graphical tool (GUI), although reduction tasks can also be completed without the GUI. We re-designed the tool to work with all platforms which are supported by IDL\footnote{{\p3d} uses widgets abundantly and can therefore not be used with GDL (cf.\ \anchor{http://gnudatalanguage.sourceforge.net}{http://gnudatalanguage.sourceforge.net}).} (version 6.2 or higher). We stress that \textsc{p3d} can be used with full functionality without an IDL license; using the IDL Virtual Machine, all required scripts are provided to use this functionality (with all UNIX-type platforms). The installation procedure consists of setting up IDL to include the files of {\p3d}, and the supplementary routines, in the path of IDL. The software, updates, documentation and tutorials can be found at the {\p3d}-project web site, \anchor{http://p3d.sourceforge.net}{http://p3d.sourceforge.net}.

All routines and their input and output parameters are documented. When {\p3d} is run using the GUI the functionality of every interactive part is described with a comment in a status indication field, and information on all activity is optionally saved to a log file. We minimized the run-time by replacing time-consuming FOR-loops with intrinsic IDL functions. All parts of the GUI can be used with screen sizes from $1024\!\times\!600$ and larger. Moreover, the program, currently, consists of about 45\,000 lines of code. Any redundancy is kept to a minimum by adhering to the ``Don't Repeat Yourself'' principle of coding \citep[][]{HuTh:99}. Our main priority is to correct algorithm issues, which lead to erroneous scientific output, as soon as possible. Avoidable issues, which are usually related to different parts of the GUI, are corrected as our time permits.

Currently {\p3d} is configured for four IFUs: the lens array (LARR) and PPAK IFUs of the PMAS instrument \citep[for both the old $2\text{k}\!\times\!4\text{k}$ CCD and the new $4\text{k}\!\times\!4\text{k}$ CCD,][]{RoKeFe.:05,KeVeRo.:06} that is mounted on the 3.5\,m telescope at Calar Alto, the VIRUS-P IFU \citep[for both bundle~1 and the newer bundle~2,][]{HiMaSm.:08} that is mounted on the 2.7\,m Harlan J.\ Smith telescope at the McDonald Observatory, and the SPIRAL IFU \citep[see e.g.][]{SmSaBr.:04} that is mounted on the 3.9\,m Anglo-Australian telescope. Support for additional instruments can be added through appropriately formatted configuration files.

\section{Components of the data reduction}\label{sec:datareduction}
{\p3d} contains a set of routines to do the following five tasks: create a master bias, trace all spectra on the detector, create a dispersion mask, create a flat field, and extract spectra in object data. We now describe these tasks in more detail.

For each task several raw data images can be combined by {\p3d} in order to increase the signal-to-noise and remove cosmic rays. The default combination method is to use a stack of at least three images, where the minimum and maximum values of every pixel are thrown away before calculating a (min/max-)average of all images. Optionally an average or a median can be used. Prescan- and overscan-regions in the raw data are only removed immediately before spectra are extracted; this must be accounted for if external tools are used to create intermediate products. Spectrum extraction can be done using either aperture extraction, our own so-called modified optimal extraction, or using a multi-profile deconvolution optimal extraction (cf.\ Sect.~\ref{sec:drspextract}). Both optimal extraction methods are able to correct for so-called cross-talk due to overlapping spectra. Currently {\p3d} does not include tasks to apply a bad pixel map, remove scattered light, subtract the sky\footnote{There are several ways to subtract the contribution of sky emission lines from object data. For the initial release of {\p3d} we have chosen not to include any sky subtraction as it is difficult to define a general procedure for it.}, perform flux calibration, compose dithered or mosaiced combined frames, or any kind of binning of spectra.

In the first step a master bias image is created by combining a set of at least three bias images. This master bias is subtracted from the raw data in all consecutive steps.

In the second step the position of every spectrum is determined along the dispersion axis with a well-illuminated calibration exposure, using a continuum lamp or twilight flats. The resulting trace mask is used in all consecutive steps when spectra are extracted. Finally, cross-dispersion profiles are calculated for all spectra and wavelengths in order to allow optimal extraction.

In the third, optional, step a dispersion correction is determined for every spectrum, using one or several arc lamp exposures. An extracted flat field image is created in the fourth, also optional, step in order to correct for wavelength-dependent variations of every spectrum and for differences in the fiber-to-fiber throughput. In the fifth, and final, step all spectra are extracted from object exposures, optionally applying first the dispersion correction and thereafter the flat-field correction.

\begin{figure}
\centering
\includegraphics{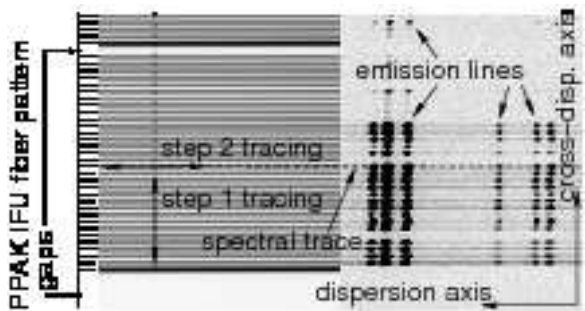}
\caption{This figure shows two adjacent PPAK raw data sections highlighting one group of 32 spectra. We show continuum lamp data on the left-hand side and object data, about H$\alpha$, on the right-hand side. Darker shades indicate stronger intensities. Horizontal tick marks, which we show on the left hand side of the image, indicate positions of individual spectra. The vertical dashed line indicates the search direction of step 1 of the tracing algorithm. The horizontal dashed line indicates, for one spectrum, the trace that results after step 2. For further details, see Sect.~\ref{sec:datareduction}.}
\label{sandinf1}
\end{figure}

In order to illustrate typical properties of IFU raw data we show a section of object data in Fig.~\ref{sandinf1}, together with the corresponding continuum lamp data, which were taken with the PPAK IFU. Every horizontal line marks the position of a spectrum, the two brightest lines (that are only seen in the left hand side image) show calibration fiber spectra of PPAK (these are not used in {\p3d}). The strong variation of emission line intensities of different spectra (on the cross-dispersion axis) is due to the spatially irregular mapping of fibers in PPAK \citep[cf.][]{KeVeRo.:06}. The two data sets that are shown next to each other illustrate their interdependence -- spectrum positions in the object data should match those in the continuum (calibration) data well in order to extract spectra properly.

\begin{table*}
\caption{General and instrument-specific parameters of the data-reduction algorithms of {\p3d}}
\label{sandint2}
\tabcolsep=5pt
\begin{tabular}{ll@{\hspace{10pt}}c@{\hspace{10pt}}cc@{\hspace{15pt}}cc@{\hspace{15pt}}l}
\hline\hline
\noalign{\smallskip}
Step & Property & \multicolumn{1}{c}{All} & \multicolumn{2}{c}{PMAS} & \multicolumn{1}{c}{VIRUS-P} & \multicolumn{1}{c}{SPIRAL} & Parameter name in {\p3d}\\
&&&\multicolumn{1}{c}{LARR}&\multicolumn{1}{c}{PPAK}&\multicolumn{1}{c}{bundle\,1/bundle\,2}&\multicolumn{1}{c}{both arms}\\[1.5pt]
\hline
\noalign{\smallskip}
\multicolumn{8}{l}{Automatic tracing algorithm, cf.~Sect.~\ref{sec:drtracing}:}\\
    & $n$                  && 256\phd\phn    &  382\phd\phn & 246\phd\phn  & 512\phd\phn\phn & \textsc{spnum}\\
1a  & $\left(n_\text{p}-1\right)\text{bin}_\lambda/2$ &&  \phn40\phd\phn    &   \phn40\phn\phd &  \phn29\phn\phd  &  \phn30\phn\phn\phd & \textsc{findwidth\_tr}\\
    & $\xi$                && \phn\phn0.9    &  \phn\phn0.9 &  \phn\phn0.9 &  \phn\phn0.95 & \textsc{cut\_tr}\\
1b  & $d\,\text{bin}_\dagger$ && \phn12.5 &  \phn\phn9.5 &  \phn\phn8.0 &  \phn\phn4.0\phn & \textsc{dist\_tr}\\
    & $\delta_\text{min}\text{bin}_\dagger$ &&  \phn\phn4.0 &  \phn\phn2.5 &  \phn\phn2.0 &  \phn\phn1.0\phn & \textsc{dmin\_tr}\\
    & $\delta_\text{max}\text{bin}_\dagger$ &&  \phn\phn5.5 &  \phn\phn4.0 &  \phn\phn4.0 &  \phn\phn2.5\phn & \textsc{dmax\_tr}\\
1c  & $\left(w_\text{cc}-1\right)\text{bin}_\dagger/2$ &\phn2\phd\phn& \multicolumn{4}{l}{\phn} & \textsc{centervar\_tr}\\ 
    & $G_\text{w}\text{bin}_\dagger$ && \phn\phn4.0 & \phn\phn4.5 & 5.0/4.2  & \phn\phn2.3\phn & \textsc{fwhm\_tr}\\
    & $n_\text{it}$ & \phn9\phd\phn & \multicolumn{4}{l}{\phn} & \textsc{niterat\_tr}\\
2   & $f$ & 10\phd\phn & \multicolumn{4}{l}{\phn} & \textsc{refinddist\_tr}\\
    & $\left(w_\text{a}-1\right)\text{bin}_\lambda/2$ &10\phd\phn&  \multicolumn{4}{l}{\phn} & \textsc{refindwidth\_tr}\\
    & $\left(w_\text{c}-1\right)/2$ &10\phd\phn&  \multicolumn{4}{l}{\phn} & \textsc{smowidth\_tr}\\
    & $\left(w_\text{d}-1\right)/2$ &\phn5\phd\phn&  \multicolumn{4}{l}{\phn} & \textsc{dispsmowidth\_tr}\\[1.0ex]
\multicolumn{8}{l}{Spectrum extraction, cf.~Sects.~\ref{sec:drspap}--\ref{sec:drspmpd}:}\\
    & $\left(x_\text{w}-1\right)\text{bin}_\dagger/2$ && \phn\phn6.0 & \phn\phn4.5 & 3.4/3.3 & \phn\phn1.5\phn & \textsc{profwidth\_ex}\\
    & $\left(\overline{x}_\text{w}-1\right)\text{bin}_\dagger/2$ && \phn\phn8.0 & \phn\phn8.0 & \phn\phn8.0 & \phn\phn5.0\phn & \textsc{profwidth\_ctex}\\
    & $n_{\text{mpd}}$ & \phn1\phd\phn & \multicolumn{4}{l}{\phn} & \textsc{mpdnprof}\\[1.0ex]
\multicolumn{8}{l}{Line profile fitting, cf.~Sects.~\ref{sec:drspmop}, \ref{sec:drspmpd}:}\\
    & $m_l$                           && \phn16\phd\phn & 31--32 & 21--44/14--29 & \phn32\phd\phn\phn & \textsc{lprofn\_tr}\\
    & $n_\text{f}\text{bin}_\lambda$ && \phn42\phd\phn &   \phn12\phd\phn   &   104\phd\phn  & \phn22\phd\phn\phn  & \textsc{lprofdint\_tr}\\[1.0ex]
\multicolumn{8}{l}{Dispersion mask creation, cf.~Sect.~\ref{sec:drwavecal}:}\\
    & $p$ & \phn4\phd\phn & \multicolumn{4}{c}{\phn} & \textsc{polynomialorder\_dm}\\
    & $\left(s_\text{w}-1\right)\text{bin}_\lambda/2$ & \phn4\phd\phn & \multicolumn{4}{l}{\phn} & \textsc{linewidth\_dm}\\[1.0ex]
\multicolumn{8}{l}{Flat fielding, cf.~Sect.~\ref{sec:drflatfield}:}\\
    & $\left(w_\text{ff}-1\right)\text{bin}_\lambda/2$ & 20\phd\phn&  \multicolumn{4}{l}{\phn} & \textsc{smowidth\_ff}\\
    & $p_\text{ff}$ & \phn7\phd\phn  & \multicolumn{4}{l}{} & \textsc{deg\_polyfit\_ff}\\[1.0ex]
\hline
\noalign{\smallskip}
\end{tabular}\\
\textbf{Notes.}\ Excepting the dimensionless properties $n$, $\xi$, $n_\text{it}$, $m_l$, $n_\text{f}$, $p$ and $p_\text{ff}$, the unit is pixels. In the third column we specify default values of general parameters. In the last column we give the name as it is specified in the instrument-specific parameter file of the program code.
\end{table*}

In the following we assign all variable parameters a symbol so that they can be found easily inside the program code. For a quick reference we collected all symbols, and their respective default value, in Table~\ref{sandint2}, the meaning of the parameters are introduced step-by-step below. The CCD readout binning parameters are, furthermore, denoted by bin$_\lambda$ (dispersion axis) and bin$_\dagger$ (cross-dispersion axis); these parameters are set to 1 or 2.

Next we describe how {\p3d} handles the separate data-reduction tasks. Our novel spectrum tracing algorithm is described in Sect.~\ref{sec:drtracing}. Thereafter we describe our approach to spectrum extraction in Sect.~\ref{sec:drspextract}, how we wavelength calibrate the data in Sect.~\ref{sec:drwavecal}, how we flat field data in Sect.~\ref{sec:drflatfield}, and how object spectra are extracted in Sect.~\ref{sec:drobjextract}.

\subsection{An automated method for finding and tracing spectra in IFU raw data}\label{sec:drtracing}
Before spectra can be extracted they must be found and their extent be traced along the dispersion axis on the detector. In order to allow accurate tracing a continuum lamp exposure should be used where spectra are clearly visible across the full wavelength range. Most IFU-instruments are, moreover, affected by flexure, which results in changing positions of spectra on the detector as the telescope moves over time; examples of instruments without such flexure are VIRUS-P and INTEGRAL. If the flexure is significant it is important to trace with a continuum lamp exposure, which is sampled close in time to the object exposure, in order to extract data from the correct region on the detector. It may also be an option to calculate one trace mask and then shift it using knowledge about shifted positions of selected lines of arc images.

We here present an automatic algorithm for spectrum tracing that involves several steps to assure that both all expected spectra of an instrument are found and traced accurately, and that no noise feature is kept as a spectrum. The method is general and should require little if no modifications once it is setup for an instrument. No assumptions are made regarding exact positions of spectra on the detector. Instead information is required about their expected number, their orientation, and their ordering. Cosmic ray hits could, if numerous, have some effect on the outcome; they should in that case be removed in advance.\\

The algorithm is split into two steps. At first all $n$ spectra are searched along the cross-dispersion (spatial) axis for one pixel, or a set of averaged pixels, on the dispersion axis. Thereafter they are traced along the dispersion axis for the remaining pixels. All instrument-specific default parameter values of this procedure are shown in Table~\ref{sandint2} for all considered IFUs. We next describe the two tracing steps separately.

\subsubsection{Step 1: finding the spectrum positions}\label{sec:trfind}
At first local maxima, representing positions of individual spectra, are located on the cross-dispersion axis; in this process a set of adjacent pixels is used on the dispersion axis, which is typically centered on the middle pixel of the CCD. It is assumed that spectra are fairly well aligned with either axis on the detector. The following condition must hold for any local maximum,
\begin{displaymath}
f\left(y_{\hat{\jmath}-1}\right)<f\left(y_{\hat{\jmath}}\right)>f\left(y_{\hat{\jmath}+1}\right),
\end{displaymath}
where $f(y)$ is the intensity distribution across the cross-dispersion axis ($y$) for every pixel $j$, and $\hat{\jmath}$ is a pixel with a local maximum. In order to reduce effects of noise, possible cosmic ray events, and variations in the detector sensitivity across its surface, the local maxima are searched in a set of $n_\text{p}$ adjacent pixels on the dispersion axis ($x$). For positively identified spectra we require that the same pixel position in $y$ is found in at least a fraction $\xi$ of the set of $n_\text{p}$ pixels in $x$. By allowing maxima to be present also in the next pixel ($y_{\hat{\jmath}+1}$) slightly tilted spectra are found as well. The statistical probability for spurious detections of noise features as spectra, which depends on the instrument setup, is nevertheless significant using only this step.

\begin{figure*}
\centering
\includegraphics{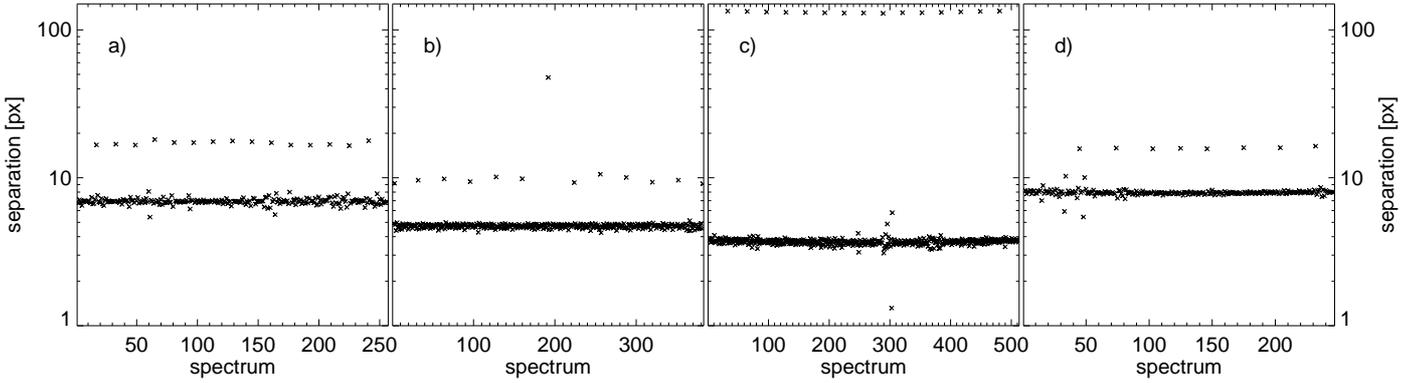}
\caption{In this figure we illustrate how adjacent spectra are separated for four IFUs: \textbf{a)} PMAS-LARR (using bin$_\dagger\!=\!2$), \textbf{b)} PMAS-PPAK (bin$_\dagger\!=\!2$), \textbf{c)} SPIRAL (bin$_\dagger\!=\!1$), and \textbf{d)} VIRUS-P (bundle 1; bin$_\dagger\!=\!1$). The ordinates are logarithmic and common to all four panels. The separation to the previous spectrum is indicated with the symbol $\times$. In panel \textbf{c} (\textbf{d}) the lowermost $\times$ indicates (the two lowermost $\times$ indicate) the calculated offset of a dead fiber. For further details, see Sect.~\ref{sec:trfind}.}
\label{sandinf2}
\end{figure*}

Next, in step 1b, knowledge of the spectrum pattern separation is used to filter out spectra from the sequence of local maxima of step 1a. The main assumption is that consecutive spectra are separated by an instrument-specific distance (pitch) $d(y)$. The first maximum in the sequence is used as a starting point of a sequence. In order to match a sequence the distance to the subsequent maximum must fulfill
$$m\!\times\!d(y)-\delta_\text{min}\le m\!\times\!d(y)\le m\!\times\!d(y)+\delta_\text{max},$$
where $m\!\times\!d$ is an integer multiple of $d$ ($m\!\ge\!1$) and $\delta_\text{min}$ and $\delta_\text{max}$ specify the permitted deviations. Mismatched maxima begin another sequence. By allowing such deviations it is possible to permit gaps between groups of spectra where the separation is non-constant. The spectrum pitch $d$, $\delta_\text{min}$ and $\delta_\text{max}$ must be determined for every instrument in advance. The spectrum separation $d(y)$ is in general constant with $y$, although with the INTEGRAL IFU it varies across the detector \citep[see e.g.][this is not yet handled by {\p3d}]{ArCaCa.:98}. In Fig.~\ref{sandinf2} we show how the separation of consecutive spectra varies for the four supported IFUs (similar figures can be used to fine-tune $d$, $\delta_\text{min}$ and $\delta_\text{max}$ for additional IFUs). Large separations indicate gaps between groups (or banks) of spectra. The separation between spectra belonging to separate groups is roughly constant for all four IFUs. Once all maxima have been traversed the longest resulting sequence is selected as the sequence of real spectra.

In step 1c a cross-correlation is made between the sequence of maxima of step 1b and a pre-defined instrumental pattern, that specifies the expected separation and number of spectra. This pattern is defined as a list of spectrum gaps, assuming spectra are separated as is described in step 1b (setting $m\!=\!1$). For example, if two groups of spectra are separated by a distance corresponding to $2\!\times\!d$ then this corresponds to one entry in the list of gaps. After the cross-correlation positions of missing spectra, due to e.g.\ dead or unused fibers, are inserted separately. Those positions, which could not be identified in steps 1a and 1b, are interpolated or extrapolated from the position of the nearest found spectrum. In this way the number of spectrum positions in the returned sequence is always as expected, viz.\ $n$ (this is a useful property when working with data of IFUs such as VIMOS that has many fibers with poor throughput). After this third step the probability that a noise feature is identified as a spectrum is negligible.

Finally, more accurate positions of the identified sequence of spectra are calculated by weighting with the cross-dispersion profile of the data. In this weighting the width of the used spectrum section is $w_\text{cc}$ pixel. The full-width at half-maximum (FWHM) of the Gaussian profile that is used in the weighting is $G_\text{w}\,$pixel. This procedure is in every case iterated $n_\text{it}$ times.

\begin{figure*}
\centering
\includegraphics{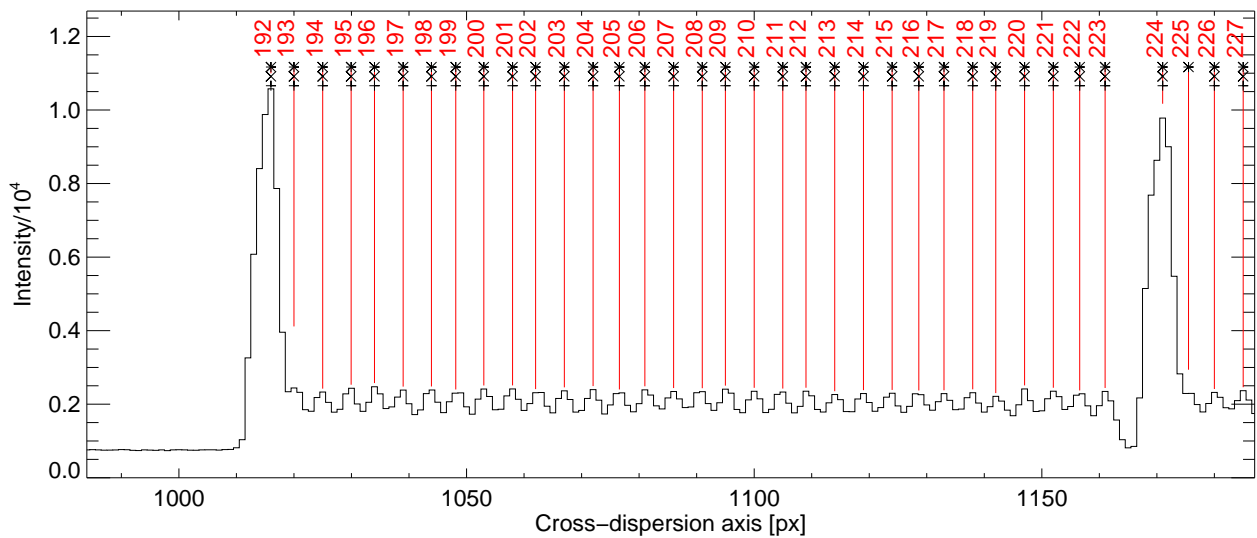}
\caption{This figure shows the result of step 1 of the tracing algorithm for continuum lamp data of the PPAK IFU, cf.\ Sect.~\ref{sec:trfind}. In this case the used CCD binning setup was $\text{bin}_\lambda\!\times\!\text{bin}_\dagger\!=\!2\!\times\!2$. The intensity is shown for the centermost pixel on the dispersion axis. Only 36 of 382 spectrum positions of PPAK are shown, every identified position is indicated with a vertical line and a spectrum number. Spectrum positions found in step 1a are indicated with a plus symbol ($+$). Positions found in step 1b are indicated with a times symbol ($\times$), and finally adopted spectrum positions are indicated with asterisk symbols (*). Note that the two spectra 192 and 224 are spectra of calibration fibers.}
\label{sandinf3}
\end{figure*}

In Fig.~\ref{sandinf3} we show an example of the outcome of step 1 of the tracing algorithm for calibration data, which were sampled with the PPAK IFU. For a direct comparison we show the same section of the spectrum that we present in Fig.~\ref{sandinf1}. Note that 35 out of 36 visible spectra were identified in steps 1a and 1b. The position of spectrum 225 is hidden in the much stronger spectrum 224. The position of this spectrum was determined in step 1c, where the spectrum pattern defined its location (as $d$ pixels distant from spectrum 224). All spectra, but those of the calibration fibers 192 and 224, are to a lesser degree affected by cross-talk, cf.\ Sect.~\ref{sec:dact}.

\subsubsection{Step 2: tracing along the dispersion axis}\label{sec:trdisp}
In a second step we calculate spectrum positions for all wavelength bins, starting with the resulting weighted positions of step 1. If spectra are traced individually along the dispersion axis there is a risk that some spectra could be lost, and because of effects of pixel subsampling an oscillating pattern of individual spectra might be found in the trace mask. The amplitude of the oscillations decreases with the profile width and becomes significant with smaller widths. The associated oscillation period depends on the angle between the spectrum and the row of pixels, and corresponds to the number of pixels in a row of pixels hosting a maximum. If positions are smoothed, using appropriately selected smoothing parameters, such oscillations can be made smaller, cf.~Sect.~\ref{sec:datm}.

We use the following approach. At first the data set is resized, by a defined factor $f$, on the dispersion axis, in order to reduce effects of noise. In this process the value at every position is averaged over $w_\text{a}\,$pixel. For every bin on the dispersion axis we then calculate weighted spectrum positions, starting with the already known position of a neighbor wavelength bin. In order to remove subpixel-sampling effects the positions are thereafter smoothed twice using a box-car of width $w_\text{c}\,$pixel, first using a median and then using an average. After all positions are calculated we smooth the positions on the dispersion axis, using a box-car of width $w_\text{d}\,$pixel.

Finally we calculate spectrum positions for all pixels on the dispersion axis. Using the smoothed and rebinned data set remaining positions are calculated by linear interpolation. Values at either end are extrapolated.\\

\subsubsection{Adjusting the automatic tracing procedure}\label{sec:trvary}
The philosophy of the described tracing algorithm is that it is automatic, and no modifications of the default values of Table~\ref{sandint2} should be required. However, in some circumstances it may turn out that the algorithm is unable to correctly identify all spectra. If this were the case the recommended procedure is to first vary $\xi$ to see if the number of spectrum misfits can be made smaller (step 1a). A second option is to vary $\delta_\text{max}$, and maybe also $\delta_\text{min}$, in order to find outliers (step 1b). If the number of misfits is large it may be necessary to add or remove one entry in the list of spectrum gaps (step 1c). If the spectrum separation parameter $d$ is changed it is in any case necessary to modify this list.

\subsection{Introducing the three spectrum extraction methods}\label{sec:drspextract}
The flux in every wavelength bin, of any spectrum on the detector, is distributed in a profile, or aperture, on the cross-dispersion axis. The shape and extent of the profile depends on the instrumental setup. If many spectra are squeezed onto the surface of the detector, as is often the case with IFSs, there is likely some overlap between profiles of adjacent spectra, resulting in cross-talk. The smaller the spectrum separation (pitch) and the larger the spectrum width are, the greater the overlap is. By selecting an accurate method of spectrum extraction it is mostly possible to separate overlapping spectra well and attain both accurate and precise values of the flux. Such an approach can, however, be very time-consuming, which is why it is worth testing if a simpler method suffices.

From here on we assume that the data set $\mathbf{o}$, that contains spectra, is bias-subtracted. The variance of the flux for every pixel $k$ in $\mathbf{o}$ is then \citep[cf.\ e.g.][]{Ho:06},
\begin{eqnarray}
V_{\text{o},k}\!=\!|\mathbf{d}_{k}-\mathbf{b}_{k}|/g_{k}+r_{k}^2+V_{\text{b},k},\quad V_{\text{b},k}\!=\!r_{k}^2/n_\text{b},
\label{eq:Vo}
\end{eqnarray}
where $\mathbf{d}$ is the raw data with spectra, $\mathbf{b}$ is the master bias, $g$ is the CCD gain ($e^{-}/\text{ADU}$; ADU is the analog-to-digital unit), $r$ is the readout noise (ADU), $V_\text{b}$ is the variance of the averaged master bias, and $n_\text{b}$ is the number of individual bias images, which were used when combining the images. A pixel index is added to the gain and the readout noise in order to account for instruments where these properties vary across the CCD surface (such as is the case with the new $4\text{k}\!\times\!4\text{k}$ CCD of PMAS that is read out in four blocks, which have to be combined before the data is used).

Next we describe the standard aperture extraction in Sect.~\ref{sec:drspap}. Thereafter we introduce our modified optimal extraction (MOX) method in Sect.~\ref{sec:drspmop}, and how we implement the multi-profile deconvolution (MPD) method in Sect.~\ref{sec:drspmpd}.

\subsubsection{Standard aperture extraction}\label{sec:drspap}
In our first approach we use an aperture of pre-defined width with a top-hat (box-car) profile that is set to 1 within the aperture, and 0 outside. This method is also referred to as tramline extraction \citep[see e.g.][hereafter {\rSB}]{ShBi:10}. We account for aperture boundaries inside of pixels by adding fractions of flux of such pixels. The integrated flux $f$ and the corresponding variance $V_\text{s}$ of every spectrum $i$ and the contributing pixels $j\,\!(i)$ of every profile, for the wavelength bin $\lambda$, are,
\begin{eqnarray}
f_{i\lambda}=\sum_{j\,\!(i)}\mathbf{d}_{j\lambda}-\mathbf{b}_{j\lambda},\quad V_{\text{s},i\lambda}=\sum_{j\,\!(i)}V_{\text{o},j\lambda}.
\label{eq:fs}
\end{eqnarray}
We present the instrument-specific default aperture widths, $x_\text{w}$, in Table~\ref{sandint2} for all four IFUs.

Although almost all flux can be collected with an aperture extraction for IFUs such as LARR this is still an inefficient method if fluxes are low. If this is the case the contribution of readout-noise from the outer pixels of an aperture can become significant, and even dominate flux errors, cf.\ Sect.~\ref{sec:daex}. Since this is the fastest method of spectrum extraction it is, nevertheless, the default method of {\p3d}.

\subsubsection{Modified optimal extraction (MOX)}\label{sec:drspmop}
Optimal extraction \citep[hereafter \rH]{Ho:86} is a more accurate method than aperture extraction. In this method line profiles are used to weight flux from separate pixels across apertures. With such line profiles it is also easier to filter out pixels, which are hit by cosmic rays. Although the process of extraction is straightforward the calculation of line profiles is demanding with IFUs. The most direct approach requires a separate profile to be calculated for every spectrum and wavelength bin. This is not only a time-consuming process, but also makes it difficult to correct for effects of noise and cross-talk; {\rS} attempts such an approach and draws a similar conclusion. The number of free parameters may be almost equal to the number of pixels, and possible effects of pixel subsampling (see above) are neglected.

Starting at one wavelength we simultaneously fit a group of line profiles, using a pre-selected function. Doing this we use the same continuum image which we used to calculate traces. The number of spectra $m_l$ in every group $l$ depends on the instrument setup. We show the ranges of $m_l$ which we use with the four IFUs in Table~\ref{sandint2}. The function can be selected to be a Gaussian function, a Lorentzian function, an approximative Voigt profile, or a double Gaussian function. As we currently ignore any broad component due to scattered light it appears sufficient to use a single Gaussian profile for all IFUs.

Since the profile width typically changes slightly across the detector it seems inappropriate to fit all spectra simultaneously, and such an approach is also computationally much more time-consuming. For every group of spectra we (by default) assume fixed profile center positions using the trace mask (the center positions can be fitted as well, if necessary). Using a Gaussian line profile the result of the fit consists of the spectrum width $\overline{G_{\text{w},l}}$, $l$ intensities, and the zero-level and the gradient of the linear fit to the background. Thereafter the same procedure is carried out for a set of additional wavelength bins, and as a last step we interpolate the profile parameters for all intermediate wavelength bins using cubic splines. The number of wavelength bins, that are used to calculate the fits, $n_\text{f}$, should be selected to allow a reasonable interpolation for intermediate wavelengths using such splines. Finally, before spectrum extraction the profile parameters are used to calculate pixel-based line profiles for every spectrum and wavelength bin. Every profile is thereafter normalized. 

In comparison to ``plain'' optimal extraction our modified approach allows a correction for overlapping spectra (i.e.\ cross-talk). Including masking of pixels, which are hit by cosmic rays, and correction for cross-talk, the equation of the modified optimally extracted flux becomes,
\begin{eqnarray}
f_{i\lambda}=\frac{\sum_{j\,\!(i)}M_{ij\lambda}P_{ij\lambda}\left(\mathbf{d}_{j\lambda}-\mathbf{b}_{j\lambda}\right){V'}_{\text{o},j\lambda}^{-1}}{\sum_{j\,\!(i)}M_{ij\lambda}P_{ij\lambda}^{2}\Gamma_{ij\lambda}V'_{\text{o},j\lambda}},
\label{eq:fo}
\end{eqnarray}
where $P$ is the normalized line profile, $M$ is a profile mask ($M\!\equiv\!1$ unless cosmic ray removal is used, see below), $\Gamma$ is a fractional profile ($0\!\le\!\Gamma\!\le\!1$, $\Gamma\!\equiv\!1$ unless a cross-talk correction is applied, see below), and $V'_{\text{o},j\lambda}\!=\!V_{\text{o},j\lambda}$. In this case the flux is integrated across $j$ using a wider aperture than with the standard extraction ($\overline{x}_\text{w}$, cf.~Sect.~\ref{sandint2}). The variance of the modified optimal spectrum estimate is, moreover,
\begin{eqnarray}
\quad V_{\text{s},i\lambda}=\frac{\sum_{j\,\!(i)}M_{ij\lambda}P_{ij\lambda}}{\sum_{j\,\!(i)}M_{ij\lambda}P_{ij\lambda}^{2}\Gamma_{ij\lambda}V'_{\text{o},j\lambda}}.
\label{eq:Vos}
\end{eqnarray}
After $f_{i\lambda}$ is calculated using ${V'}_{\text{o},j\lambda}\!=\!V_{\text{o},j\lambda}$ (Eq.~\ref{eq:Vo}) we instead use
\begin{eqnarray}
{V'}_{\text{o},j\lambda}=\left|f_{i\lambda}P_{ij\lambda}\right|/g_{j\lambda}+r_{j\lambda}^2
\end{eqnarray}
and iterate the solution $n_2$ times. Again, following {\rH} the default is $n_2\!=\!1$.

Next we describe how we remove cosmic rays and correct the integrated flux for cross-talk. Both methods require an inspection of the resulting outcome, in order to see that it is satisfactory. These are therefore options that must be switched on separately in {\p3d}.

\paragraph{Removal of cosmic ray hits}
In order to remove cosmic ray hits we follow the approach of {\rH} and first iterate the integrated flux (Eq.~\ref{eq:fo}) at most $n_{\text{CR}}$ times. In each iteration we mask at most one pixel ($j$) with the highest value which satisfies
\begin{eqnarray*}
M_{ij\lambda}\left\{\left(\mathbf{d}_{j\lambda}-\mathbf{b}_{j\lambda}-f_{i\lambda}\right)^2-\sigma_\text{CR}^2V'_{\text{o},j\lambda}\right\}>0,
\end{eqnarray*}
by setting $M_{ij\lambda}=0$. Here $\sigma_{\text{CR}}$ is a threshold that defines how large the deviation of one single pixel must be to be classified as a cosmic ray hit. In comparison to {\rH} we found that it is necessary to use values larger than $\sigma_\text{CR}\!=\!5$. This method sometimes has difficulties removing cosmic rays if emission lines are very strong. Unless $\sigma_\text{CR}$ is set high enough ($\ga\!10$) pixels in the line center of profiles with high intensity may be removed as cosmic rays; resulting in a strong decrease of the integrated flux, as the lost flux is not replaced with any interpolated (or expected) flux. The default value on the number of iterations is $n_\text{CR}\!=\!2$.

\paragraph{Correcting extracted flux when there is overlap between nearby spectra on the CCD}
In order to correct for cross-talk we iterate the spectrum extraction. In every iteration we first calculate a total profile across all $m_l$ spectra of group $l$. Thereafter we calculate a fractional profile $\Gamma_{ij\lambda}$ that for every contributing pixel $j\,\!(i)$ indicates which fraction, of the flux $\left(\mathbf{d}_{j\lambda}-\mathbf{b}_{j\lambda}\right)$ and the variance ${V'}_{\text{o},j\lambda}$, belongs to line profile $i$. The spectrum extraction is iterated at most $n_\text{CT}$ times, or until the maximum relative change of the calculated extracted flux of all spectra in group $l$, of two consecutive iterations, is $<\!\sigma_{\text{CT}}$. The default number of iterations and value of the threshold are $n_\text{CT}\!=\!15$ and $\sigma_{\text{CT}}\!=\!10^{-5}$. The number of required iterations depends on the data and the instrument, but it appears that less than five iterations are required, typically, using data of the PPAK IFU.

\subsubsection{Multi-profile deconvolution optimal extraction (MPD)}\label{sec:drspmpd}
In addition to the modified optimal extraction {\p3d} can also use the multi-profile deconvolution method of {\rSB}. In comparison to our method above all line intensities at one wavelength are here solved for simultaneously. Specifically, for every group of spectra the method is to minimize the residual
\begin{eqnarray}
R_{\text{f},\lambda}=\frac{1}{2}\sum_{j\,\!(i)}\frac{\left(\mathbf{d}_{j\lambda}-\mathbf{b}_{j\lambda}-\sum_if_{i\lambda}P_{ij\lambda}\right)^2}{V_{\text{o},j\lambda}}.
\end{eqnarray}
Assuming $\partial R_{\text{f},\lambda}/\partial f_{i\lambda}\!=\!0$ we then find and solve
\begin{eqnarray}
\sum_if_{i\lambda}c_{\text{f},il\lambda}=b_{l\lambda}
\label{eq:mpdone}
\end{eqnarray}
for the intensities $f_{i\lambda}$, where
\begin{eqnarray*}
c_{\text{f},il\lambda}=\sum_{j\,\!(i)}\frac{P_{ij\lambda}P_{lj\lambda}}{V_{\text{o},j\lambda}}\quad\text{and}\quad b_{\text{f},l\lambda}=\sum_{j\,\!(i)}\frac{\left(\mathbf{d}_{j\lambda}-\mathbf{b}_{j\lambda}\right)P_{lj\lambda}}{V_{\text{o},j\lambda}}.
\end{eqnarray*}

Considering the variance we calculate it using a similar approach by minimizing the residual
\begin{eqnarray}
R_{\text{V},\lambda}=\frac{1}{2}\sum_{j\,\!(i)}\left(V_{\text{o},j\lambda}-r_{j\lambda}^2-\sum_iV_{\text{s},i\lambda}P_{ij\lambda}\right)^2\!\!\!.
\end{eqnarray}
Assuming $\partial R_{\text{V},\lambda}/\partial V_{\text{s},i\lambda}\!=\!0$ we then find and solve
\begin{eqnarray}
\sum_iV_{\text{s},i\lambda}c_{\text{V},il\lambda}=b_{\text{V},l\lambda}
\label{eq:mpdtwo}
\end{eqnarray}
for the variances $V_{\text{s},i\lambda}$, where
\begin{eqnarray*}
c_{\text{V},il\lambda}=\sum_{j\,\!(i)}P_{ij\lambda}P_{lj\lambda}\quad\text{and}\quad b_{\text{V},l\lambda}=\sum_{j\,\!(i)}\left(V_{\text{o},j\lambda}-r_{ij\lambda}^2\right)P_{il\lambda}.
\end{eqnarray*}

The other two spectrum extraction methods use a relatively limited size of the aperture. In this method we use the aperture width
\begin{eqnarray*}
\hat{x}_\text{w}=2n_\text{mpd}d/\text{bin}_\dagger+1,
\end{eqnarray*}
where $n_\text{mpd}$ is the number of neighbor profiles, on either side of every spectrum, which are considered. By default $n_\text{mpd}\!=\!1$, but it can be increased to include more spectra if line profiles are broad. The profiles $P$ are calculated according to the description in Sect.~\ref{sec:drspmop}.

Following the approach of {\rSB} the systems of equations (Eqs.~\ref{eq:mpdone} and \ref{eq:mpdtwo}) are solved for $f_{i\lambda}$ using a tri-diagonal solver if $n_\text{mpd}\!=\!1$. For $n_\text{mpd}\!\ge\!1$ there is a choice of solving the equations using either a solver for a sparse diagonal matrix or singular value decomposition.

\subsection{Preparing a dispersion solution}\label{sec:drwavecal}
Depending on the optical path through instrument and fibers resulting spectra of separate spatial elements of an IFU are shifted and stretched relative to each other. An image with extracted spectra of an arc lamp exposure shows emission lines as curves across the cross-dispersion axis. In order to find a dispersion solution, which is unique to an IFS exposure, all spectra should be aligned and stretched to use the same starting wavelength and size of wavelength bins.

In order to calculate a dispersion mask {\p3d} requires a list of lines with known wavelengths, some information about the expected wavelength range of the used setup, and an arc lamp image. In brief, the list of lines is first modified to match pixel positions of corresponding entries in the data. Thereafter a polynomial is fitted between pixel positions and corresponding wavelengths. Next we describe the individual steps in more detail.

In a first step a list of lines, with well known wavelengths, is selected; either automatically using information in the arc image file header, or as defined by the user. For PMAS, SPIRAL, and other instruments where the canonical reflective grating equation applies, an initial estimate of the wavelength range is calculated using information about the instrumentation setup in the data header. For VIRUS-P the wavelength range is pre-defined as 3620--5910\,{\AA}\footnote{This wavelength range can, if necessary, easily be changed by providing {\p3d} with alternative lower and upper values.}.

In a second step a line mask curvature is determined by calculating positions for one or two lines in the arc frame. The algorithm to do this involves locating the maximum position of one line in one spectrum, and then track this maximum through all other spectra. Emission lines of separate fibers in PMAS data using the $2\text{k}\!\times\!4\text{k}$-CCD are, to first order, only shifted in wavelength relative to each other. In this case we found that it suffices to select one emission line in the data in order to calculate the curvature accurately enough. VIRUS-P, SPIRAL, and PMAS using the $4\text{k}\!\times\!4\text{k}$-CCD are different, for these instruments the dispersion varies across the IFU surface. In this case we selected two clearly separated emission lines in the data, in order to calculate both the curvature and the change of dispersion.

\begin{figure*}
\centering
\includegraphics{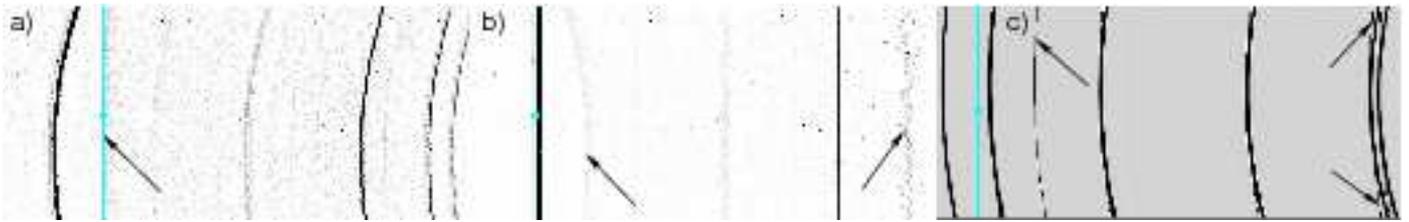}
\caption{This figure shows three extracted arc frames of: \textbf{a)} LARR, \textbf{b)} LARR after the dispersion solution was applied, and \textbf{c)} VIRUS-P. Colors are modified to enhance various features. In each panel the vertical line with a disk indicates a preset reference spectrum and wavelength bin. In panel \textbf{a} the arrow points at the merging point of a shorter (1s; left part) and a longer (10s; right part) exposure, note that this straight line is curved in panel \textbf{b} (left arrow). In panel \textbf{b} the right arrow points at an emission line that was outside the calibration range. In panel \textbf{c} the left arrow indicates the curved shape of a line in the line mask, after the line geometry has been calculated. The two right arrows indicate a mismatch between the line mask and the data for redder wavelengths. For further details see Sect.~\ref{sec:drwavecal}.}
\label{sandinf4}
\end{figure*}

We present examples of curved arc lines, in the blue wavelength range, for LARR in Fig.~\ref{sandinf4}a (for lines of mercury) and for VIRUS-P (bundle 2) in Fig.~\ref{sandinf4}c (for lines of cadmium and mercury). In order to illustrate the change of dispersion of VIRUS-P we, in this case, calculated the curvature using one line, instead of two. As Fig.~\ref{sandinf4}c shows the curvatures of lines in the line list badly matches the data in the redder part of the image. If two lines are used instead to calculate the curvature the match is excellent across all spectra.

In a third step the line mask is shifted along the dispersion axis in order to achieve a rough match between entries of the line mask and lines in the arc image. If lines cannot be matched across the entire dispersion axis the constant pre-estimated dispersion can be changed manually. With the PMAS $4\text{k}\!\times\!4\text{k}$-CCD, in particular, the dispersion changes between the blue and the red ends of the detector. In this case it is necessary to fit a preliminary non-linear dispersion solution to an interactively selected set of lines. If the selected line mask, moreover, contains saturated lines, or if there are more entries in the input line mask than are visible in the data, such entries can be removed interactively in a fourth step.

More precise pixel positions of entries in the line mask are then calculated by correlating every line in the line mask with the data, this is done for every spectrum. In this fifth step more precise pixel positions are by default calculated using a Gaussian function fitting. Alternatively they can be determined using a much faster iterated average-weighting scheme. The region that is searched for an intensity maximum normally spans a range of $s_\text{w}\,$pixel on the dispersion axis. If emission lines are too tightly packed, in either the line mask or in the data, this step may fail; in this case it is recommended to remove a few entries in the line mask.

The sixth, and final, step is where a dispersion mask is created. A (linear) polynomial of pre-defined order $p$ is fitted to the pixel positions at all wavelengths, for every spectrum, in the line mask. The polynomial order can be set to any value, the default is to use a low order with $p\!=\!4$. Residuals of the fits are stored to the data reduction log file, and can also be viewed interactively. The dispersion mask is saved as $p+1$ fitting parameters of every spectrum.

Finally, we show an example of how an extracted arc lamp frame appears before and after the dispersion solution is applied in Figs.~\ref{sandinf4}a and \ref{sandinf4}b. Note the extremely noisy line in the right part of Fig.~\ref{sandinf4}b. This line lies outside the range of selected lines, which were used in the creation of the dispersion mask, and was therefore inadequately calibrated.

\subsection{Flat fielding the extracted IFU data}\label{sec:drflatfield}
The optical path and transmission efficiency of individual spatial elements across an IFU typically vary. Differences appear both as wavelength-dependent variations and as a variance in the fiber-to-fiber throughput. For multiple-detector instruments, such as VIMOS, there is likely also a difference in the detector-to-detector throughput. After spectra are extracted and wavelength calibrated {\p3d} can correct for these variations by normalizing the data with an extracted flat field image, which combines the required corrections.

A correction for a variable fiber-to-fiber throughput is applied by dividing every spectrum of a flat field image with its mean spectrum. Correcting for wavelength-dependent variations every spectrum is at first smoothed across the dispersion axis using a box-car of width $w_\text{ff}\,$pixel. Thereafter it is replaced with a (linear) polynomial fit of order $p_\text{ff}$. Finally, the flat field is normalized with the mean value of all elements. Only non-zero elements are used with these operations. This smoothing minimizes the amount of noise that is added to the flat fielded spectra \citep[see e.g.][]{TBe:02,RoKeFe.:05}. This smoothing can, if required, be switched off (by setting $w_\text{ff}\!=\!0$ and $p_\text{ff}\!=\!0$).

Before a normalized flat field is created {\p3d}, by default, first calculates a trace mask for this task using the same image. If twilight exposures of flexure-affected instruments are used as flat fields this is an important aspect as it assures that the proper traces are used. The extracted flat field can, if required, be wavelength calibrated using a separate dispersion mask.

\subsection{Object extraction}\label{sec:drobjextract}
The only prerequisite to extract spectra of object data is a trace mask. Cosmic ray hits can, currently, be removed either by allowing {\p3d} to combine a set of raw data images, or by using the cosmic ray removal option of the modified optimal extraction procedure for separate images. Alternatively, for separate images they could be removed in advance, outside {\p3d}, using, for example, the approach of \citet[who calls his routine L.A.\ Cosmic]{vDo:01} or \citet{Py:04}. In principle a cosmic-ray removal option can also be added to the multi-profile deconvolution extraction procedure. We do not recommend to use this approach, however, since it is difficult to remove such hits efficiently and also keep all flux. Instead, for a future version of {\p3d} we propose to use a mask (calculated using e.g.\ L.A.\ Cosmic) that indicates cosmic-ray affected pixels. The intensity error of the masked pixels are set to a high value before the extraction, and will thereby be given a minimal weight using either of the optimal extraction routines. Errors are, furthermore, calculated and stored in a separate file for all intermediate products, but only if a master bias is specified.

After their extraction spectra are wavelength calibrated if a dispersion mask is specified. If the object data contains sky emission lines {\p3d} optionally calculates a shift of the dispersion mask based on the known wavelengths of those lines. The offset is calculated by first fitting all present sky emission lines in all spectra with a Gaussian profile. The offset is then taken as the median of the difference between the center positions and the expected wavelengths. The error of wavelength calibrated spectra should, moreover, be considered a lower limit since pixel values are not cross-correlated when interpolating the dispersion solution to a common base wavelength.

{\p3d} stores all data in row-stacked-spectra (RSS) format, although final spectra can optionally also be saved in the E3D-format \citep[see e.g.][]{KiCoFe.:04}. When the RSS-formatted file is used to view or analyze the data further, outside {\p3d}, it is necessary to use a separate table specifying the positions of the spatial elements. The tables of the {\p3d}-distribution in this case provide all necessary information. If the data need to be corrected for effects of differential atmospheric refraction we recommended to make this correction before the data is flux calibrated. The approach of \citet[cf.\ e.g.\ \citealt{SaScRo.:08}]{Fi:82} provides the most straightforward approach as it only requires information about observing conditions (the required information is mostly found in the data header).

\section{Program validation}\label{sec:dataanalysis}
In this section we present the outcome of our tests of {\p3d}. Primarily we used simulated data since the outcome then is known. In our study of properties of the trace mask and the dispersion mask we also compared outcome of {\p3d} with corresponding outcome of {\iraf} for data of the LARR IFU. A description of the observational setup for used data, in this case, can be found in \citet{ReMoVi.:09}. The {\iraf} data-reduction follows the scheme that is presented by \citet{AlGaMo.:09}, with some minor modifications \citep[see][for details]{ReMoVi.:09}.

At first we discuss the accuracy of the trace mask in Sect.~\ref{sec:datm}, and study the importance of using a cross-talk correction in Sect.~\ref{sec:dact}. Thereafter we compare the outcome of the wavelength calibration of {\p3d} with outcome which is created using {\iraf} instead in Sect.~\ref{sec:dadm}. In Sect.~\ref{sec:daex} we evaluate the accuracy of the three spectrum extraction methods. Finally, in Sect.~\ref{sec:daoo} we compare resulting spectra using the different spectrum extraction with observations of planetary nebulae.

\subsection{Estimating the accuracy of calculated trace masks}\label{sec:datm}
Accurately determined traces, i.e.\ spectrum center positions, are necessary in order to extract spectra properly (cf.~Sect.~\ref{sec:daex}). The automatic tracing algorithm of {\p3d} is mostly able to locate all spectra at one wavelength, without any interaction, in the first step. During the spectrum tracing across all wavelengths, of the second step, the accuracy of the result depends on how well the profile center positions are determined.

\begin{figure*}
\centering
\includegraphics{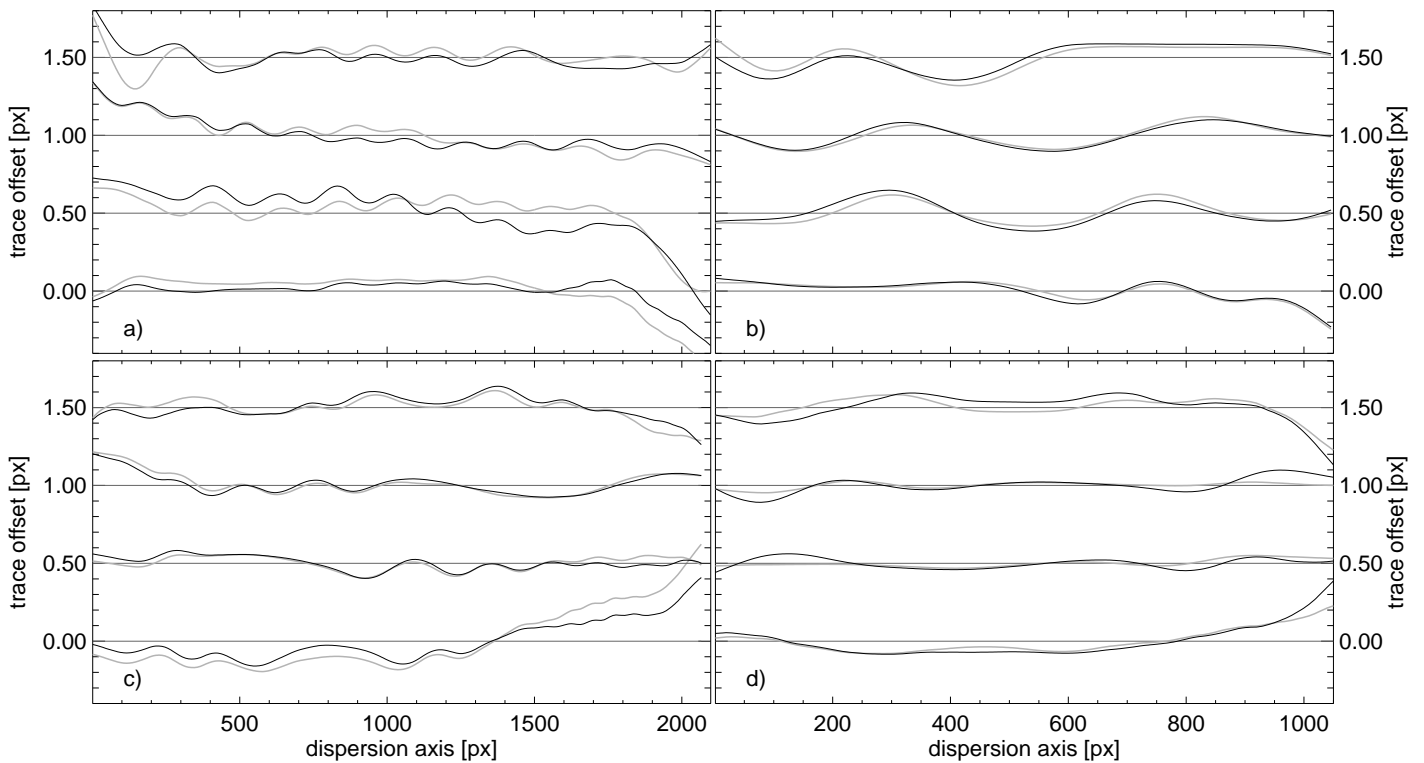}
\caption{This figure illustrates the influence of the cross-spectrum smoothing of step 2 of the tracing algorithm. Four trace residuals, which are calculated from typical continuum images, are shown for each IFU as a function of the dispersion axis: \textbf{a)} LARR, \textbf{b)} PPAK, \textbf{c)} SPIRAL and \textbf{d)} VIRUS-P; the traces are selected uniformly across the IFU surface. The black (light gray) solid lines show the respective trace where the $w_{\text{a}}\!\!=\!\!11$-smoothed ($w_{\text{a}}\!\!=\!\!21$-smoothed) trace mask is subtracted from the non-smoothed mask ($w_{\text{a}}\!=\!1$). In order to plot all four traces using the same axis they are offset by half a pixel from each other. Dark gray horizontal lines are guides. For further details see Sect.~\ref{sec:datm}.}
\label{sandinf5}
\end{figure*}

\begin{figure}
\centering
\includegraphics{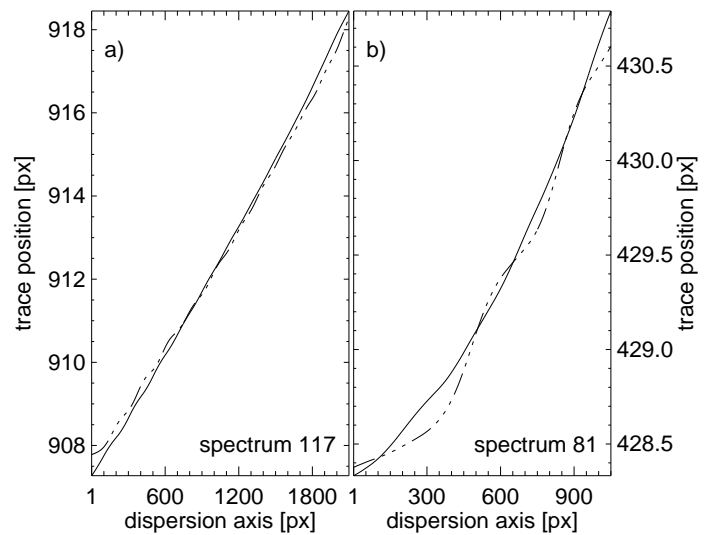}
\caption{One trace is shown for the LARR IFU in panel \textbf{a}, and for the PPAK IFU in panel \textbf{b}. The position on the cross-dispersion axis is in both cases drawn as a function of the position on the dispersion axis. The solid line shows the smoothed trace and the dash-triple-dotted line the non-smoothed trace. Compare with Figs.~\ref{sandinf5}a \& \ref{sandinf5}b, cf.~Sect.~\ref{sec:datm}.}
\label{sandinf6}
\end{figure}

In Fig.~\ref{sandinf5} we compare calculated traces, which are smoothed across several spectra ($w_\text{a}\!=\!11,\,21$, cf.\ Sect.~\ref{sec:trdisp}), with non-smoothed traces ($w_\text{a}\!=\!1$). The difference between the two sets are $\la\!0.2$ pixel in every case, except at the blue and red ends where differences are up to about 0.5\,pixel. All four IFUs show semi-periodic offsets in the non-smoothed traces, that is due to inaccurate pixel subsampling. The oscillation period of a trace corresponds to the number of illuminated pixels of a row on the detector. We illustrate this for one typical trace of the LARR IFU in Fig.~\ref{sandinf6}a, and for one similarly typical trace of the PPAK IFU in Fig.~\ref{sandinf6}b. The trace of the LARR IFU extends across about 10.5 pixels on the cross-dispersion axis, corresponding to 10.5 oscillation periods, which are seen in the second trace from the top in Fig.~\ref{sandinf5}a. Likewise the PPAK trace extends across about 2 pixels, that equals the number of oscillation periods of the two middle traces in Fig.~\ref{sandinf5}b. In addition to offsets of inaccurate pixel subsampling an additional offset is caused by the smoothing across several spectra. In this context the exact value of $w_\text{a}$ seems unimportant as differences are both lower and higher when the traces of either smoothed mask are compared with the non-smoothed mask.

\begin{figure*}
\centering
\includegraphics{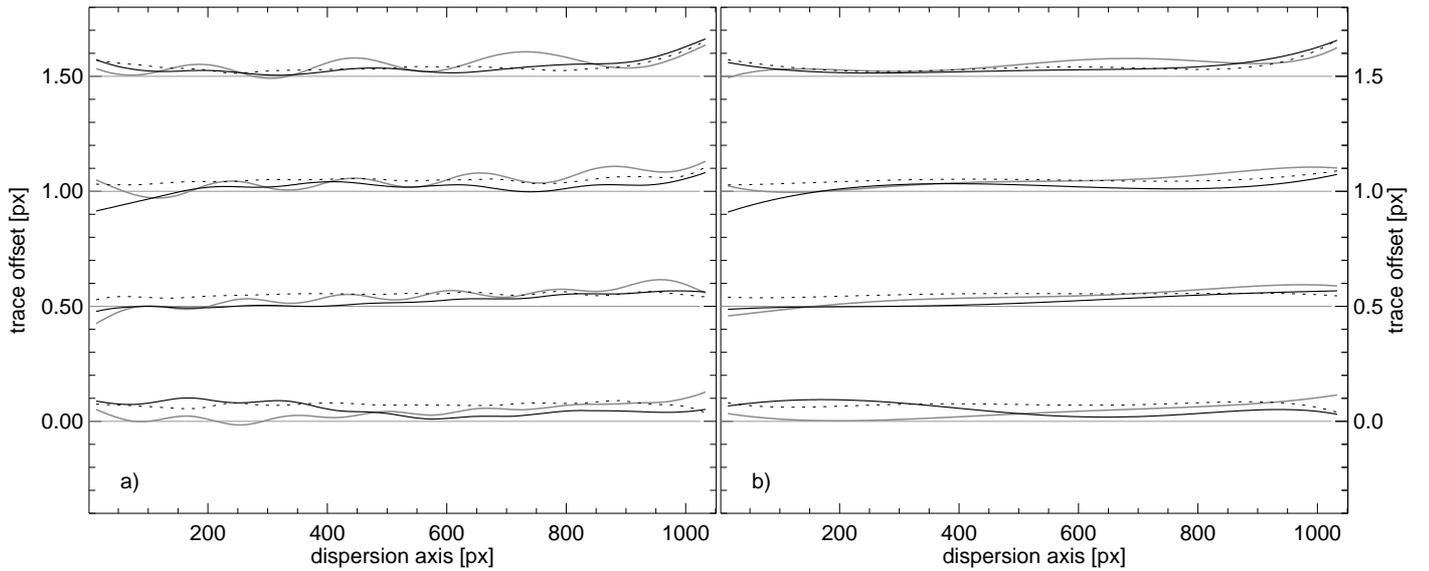}
\caption{Four trace residuals are shown for a typical continuum image of the LARR IFU as a function of the dispersion axis: \textbf{a)} trace(\iraf)-trace(\p3d), \textbf{b)} trace(\iraf)-fit(trace(\p3d)); the traces are selected uniformly across the IFU surface. The black (gray) lines show the residuals using the smoothed (non-smoothed) traces of {\p3d}. The dotted line shows residuals of traces that were fitted anew during the profile calculation. In order to plot all four traces using the same axis they are offset by half a pixel from each other. The dark gray horizontal lines are guides. For further details see Sect.~\ref{sec:datm}.}
\label{sandinf7}
\end{figure*}

\begin{figure*}
\centering
\includegraphics{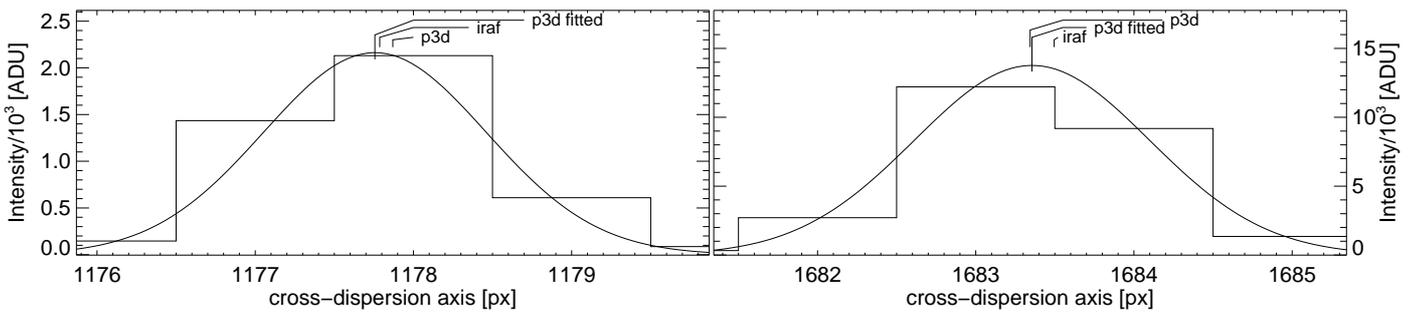}
\caption{Profile center positions are shown across the cross-dispersion axis for {\iraf}, {\p3d}, and fitted profiles of {\p3d} for two wavelength bins of two different spectra on the detector. The raw data is drawn as a histogram, and the Gaussian fitted profile is shown as a continuous curve. For further details see Sect.~\ref{sec:datm}.}
\label{sandinf8}
\end{figure*}

\begin{figure*}
\centering
\includegraphics[width=15cm]{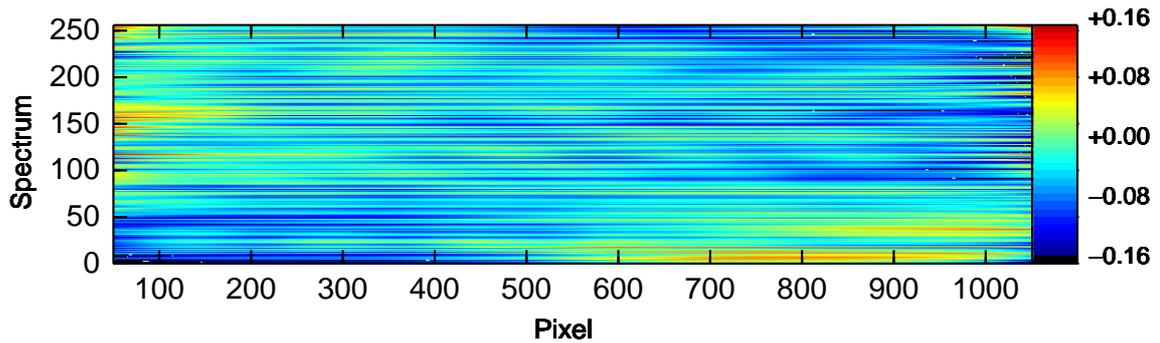}
\caption{In this image we show the difference between a trace mask that is calculated using {\p3d} and a corresponding trace mask that is calculated using {\iraf}. For further details see Sect.~\ref{sec:datm}.}
\label{sandinf9}
\end{figure*}

Next we compare traces created with {\p3d} and {\iraf} for the LARR IFU. At first we show four residual traces where the trace mask of {\p3d} was subtracted from the trace mask of {\iraf}, cf.~Fig.~\ref{sandinf7}a. As the figure shows these residuals are always $<\!0.2\,$pixel, and mostly $<\!0.1\,$pixel. The residuals of the non-smoothed trace mask show slightly higher variations. We also show residuals using fitted traces (dotted line; where the line center positions were fitted together with the intensities when calculating the profiles). In this case the fitted traces differ slightly from both the fixed traces and those of \iraf. Note that the fitted traces do not oscillate across the dispersion axis. In comparison to the outcome of the Gaussian weighting (solid lines in Fig.~\ref{sandinf7}) this indicates a negligible influence on Gaussian-fitted center positions due to pixel subsampling.

For this test we also fitted the traces of {\p3d} with a fourth order linear polynomial, which we then subtracted from the trace mask of {\iraf}. We show the result in Fig.~\ref{sandinf7}b. In this case the non-smoothed traces show a better agreement with the traces of {\iraf}. As both panels in the figure show the difference is minute when the fitted traces are replaced with a polynomial fit. The minimum and maximum values of the residual of all spectra are -$0.17$ and $0.22\,$pixel. The mean and standard deviation are -$0.043$ and $0.044\,$pixel. In Fig.~\ref{sandinf9} we show an image of the residuals of all spectra. The discrepancy is, again, the largest at the blue and the red ends.

In order to compare the calculated center positions with the raw data we indicate the center positions of each approach in Fig.~\ref{sandinf8}, for two wavelengths. We selected the bluemost part of the second trace from the top in Fig.~\ref{sandinf7} (lefthand side panel), and the redmost part of the topmost trace (righthand side panel), since this is where the difference is the largest in the shown residuals. In the former case the fitted center position of {\p3d} lies closer to the position of \iraf, the difference compared with the pre-calculated trace mask position is 0.11\,pixel. In the latter case the same difference is only 0.012\,pixel, while the position of {\iraf} is more offset. The profile center positions of {\p3d} are here more accurate than those of {\iraf}.

We conclude that the error of (fixed) traces, which are calculated using {\p3d}, should be $\la\!0.2$\,pixel. For spectra, which are located close to the center of the detector, this value is likely better. At the blue end of the detector, for $\lambda\!\la\!4000\,${\AA} the accuracy is likely worse due to a lower detector sensitivity. If high accuracy is required, as may be the case when cross-talk is present (cf.\ Sect.~\ref{sec:daex}), the preferred method is to replace the fixed center positions of the trace mask with fitted positions. Note, however, that it may be difficult to achieve the highest accuracy of the spectrum extraction with flexure-affected instruments. Because, spectra might have moved on the detector during the time between the calibration exposure and the object exposure. A solution to this problem could be to re-center the calculated profiles using the object data.

\begin{figure*}
\centering
\includegraphics{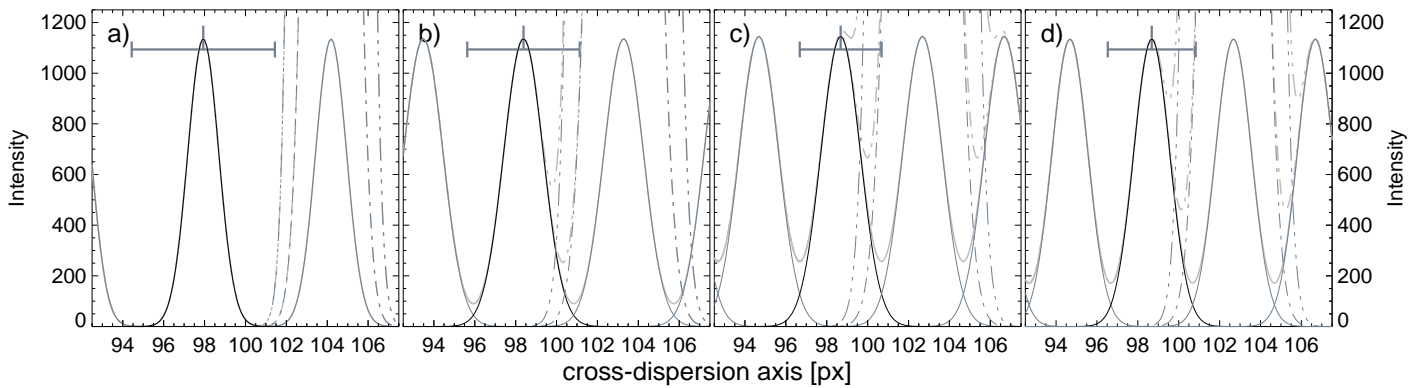}
\caption{This figure illustrates effects of fiber-to-fiber cross-talk between neighbor spectra for typical idealized profiles of four IFUs: \textbf{a)} LARR, \textbf{b)} PPAK, \textbf{c)} SPIRAL (Blue arm), and \textbf{d)} VIRUS-P (bundle~2; bin$_\dagger\!=\!2$). The abscissa shows a part of the cross-dispersion axis, while the ordinate shows the intensity in an arbitrary unit. The reference profile is drawn with a black solid line, and neighbor profiles with gray lines. In every case profiles, which are drawn with solid lines, have equal intensity. Profiles that are drawn with a dash-dotted (dash-dot-dot-dotted) line are 10 (100) times as intense. The summed spectrum is drawn with a thick light gray line. The center position and width of the aperture are indicated at the top of each panel. For further details compare with the values in Table~\ref{sandint3} and also see Sect.~\ref{sec:dact}.}
\label{sandinf10}
\end{figure*}

\begin{table*}[t]
\caption{Parameters and outcome of our idealized study of the influence of cross-talk}
\label{sandint3}
\begin{tabular}{lccccccccc}
\hline\hline
\noalign{\smallskip}
IFU & \multicolumn{1}{c}{bin$_{{\dagger}}$} & \multicolumn{1}{c}{$G_{\text{in}}$} & \multicolumn{1}{c}{$G_{\text{mean}}$}& \multicolumn{1}{c}{$G_{\text{range}}$} & \multicolumn{1}{c}{$x_\text{w}'/\text{bin}_\dagger$} & \multicolumn{1}{c}{$\delta_{\text{I}}$} & \multicolumn{1}{c}{$\delta_{\text{1/\text{R}}}$} & \multicolumn{1}{c}{$\delta_{\text{10/\text{R}}}$} & \multicolumn{1}{c}{$\delta_{\text{100/\text{R}}}$}\\
&& \multicolumn{1}{c}{$\left[\text{px}\right]$} & \multicolumn{1}{c}{$\left[\text{px}\right]$} & \multicolumn{1}{c}{$\left[\text{px}\right]$} & \multicolumn{1}{c}{$\left[\text{px}\right]$} & \multicolumn{1}{c}{$\left[\%\right]$} & \multicolumn{1}{c}{$\left[\%\right]$} & \multicolumn{1}{c}{$\left[\%\right]$} & \multicolumn{1}{c}{$\left[\%\right]$}\\[1.5pt]
\hline
\noalign{\smallskip}
LARR/V600/blue   &2& 2.00& $1.78\pm0.06$ & 1.6--2.1 &7.0&0.0004&0.03&\phn0.13&\phn\phn1.1\\
PPAK/V600        &2& 2.25& $2.27\pm0.09$ & 2.1--2.5 &5.5&0.44\phn\phn  &2.6\phn &11\phd\phn\phn &110\phd\phn\\
SPIRAL/Blue Arm$^\text{a}$ &1& 2.30& $2.25\pm0.08$ & 2.1--2.5 &4.0&3.7\phn\phn\phn &3.8\phn&19\phd\phn\phn &170\phd\phn\\
SPIRAL/Red Arm   &1& 2.30& $2.43\pm0.15$ & 2.1--2.6 &4.0&5.3\phn\phn\phn   &5.6\phn &28\phd\phn\phn &260\phd\phn\\
VIRUS-P/bundle~1 &1& 5.00& $4.73\pm0.47$ & 3.8--5.4 &7.8&5.2\phn\phn\phn   &4.4\phn &23\phd\phn\phn &200\phd\phn\\
VIRUS-P/bundle~2 &1& 4.20& $4.01\pm0.23$ & 3.7--4.5 &7.6&2.6\phn\phn\phn   &1.4\phn &\phn7.2\phn&\phn67\phd\phn\\
VIRUS-P/bundle~2 &2& 2.10& $2.07\pm0.10$ & 2.0--2.3 &4.3&1.5\phn\phn\phn   &3.6\phn &18\phd\phn\phn &160\phd\phn\\[1.0ex]
\hline
\noalign{\smallskip}
\end{tabular}\\
{$^\text{a}$ Due to low intensities in the bluemost part we only used the redmost 1650 pixels on the dispersion axis.}
\end{table*}

\subsection{On the importance of correcting for fiber-to-fiber cross-talk}\label{sec:dact}
The same calibration data that is used to calculate the trace mask is also used to calculate cross-dispersion line profiles, which are used in the optimal extraction. For most instruments there is some overlap between spectra on the cross-dispersion axis. The magnitude of the overlap depends on three factors: the spectrum separation ($d$), the width of the profiles, and the intensity of every spectrum. In order to estimate the influence of cross-talk we first calculated line profiles for different setups of each supported IFU. In this idealized study we assumed a Gaussian shape of the profile in every case. Note that this simplified treatment neglects the effect of extended profile wings, which are due to scattered light; typically at $<\!1\%$ peak intensity, but large FWHM \citep[cf.\ e.g.][]{TBe:02}.

Using data from different observing runs we first measured both the variation and the average value of the spectrum width across the IFU. We present the outcome in Table~\ref{sandint3}. The initial FWHM of the profiles of every group ($G_{\text{in},l}\!=\!G_{\text{w}}/\text{bin}_{\dagger}$) was taken from Table~\ref{sandint2}. The calculated average width and its standard deviation are given in Col.~4 ($G_{\text{mean}}$), and the full range of measured widths in Col.~5 ($G_{\text{range}}$). Note that the width varies across the detector for all IFSs, in most cases across both axes. This variation should be kept in mind when interpreting the percentages we present next.

For each IFU configuration we then calculated a set of 3--4 line profiles using the spectrum width $G_{\text{mean}}$ and the spectrum separation $d$. We integrated the flux for one (reference) profile across an aperture of pre-defined width $x_\text{w}'$ ($x_\text{w}'\!=\!x_\text{w}/\text{bin}_\dagger$, see Table~\ref{sandint2}) and calculated the fraction of the flux which fell outside the aperture ($\delta_\text{I}$). Additionally, we calculated the fraction of increased flux inside the aperture, due to the two neighbor profiles, in order to estimate the cross-talk contribution. Doing this we assumed that the intensity of the right-hand-side profile is 1, 10 and 100 times higher ($\delta_{1/\text{R}}$, $\delta_{10/\text{R}}$, $\delta_{100/\text{R}}$) than the reference profile. The outcome for each IFU setup is given in Table~\ref{sandint3}. We additionally illustrate the profiles of four setups in Fig.~\ref{sandinf10}.

This study reveals several important results. For the LARR IFU it is evident that a cross-talk correction is unnecessary. Even if there is a strong intensity gradient across the IFU, with an intensity ratio of 100 between two neighbor spectra, the amount of cross-talk is only about 1\%. The amount of flux outside the aperture of the reference spectrum of the PPAK/IFU is negligible (0.44\%). Since fibers are more tightly packed, however, the amount of cross-talk is significant. If intensity gradients are moderate (and ratios are $<\!10$) the intensity of the reference profile is increased by about 12\% due to cross-talk. With the remaining IFU configurations the amount of flux outside the aperture is 1--5\%, while the cross-talk contribution to the reference profile increases to 7--28\% for moderate overlap, and to 67--260\% for strongly overlapping profiles (intensity ratio of 100). With the exception of the LARR IFU it is clear that it is necessary to correct for cross-talk in order to achieve any level of accuracy in integrated fluxes of weaker regions of the object. Although, even for the LARR IFU an optimal extraction could be advantageous if it is important to extract weak lines accurately (cf.\ Sect.~\ref{sec:daoo}).

\citet{AlCo:98} argue that a cross-talk correction is unneccessary when neighboring spectra on the detector correspond to well-sampled neighboring positions on the sky, as is the case with the LARR and the SPIRAL IFUs (and also the VIMOS and the FLAMES-ARGUS IFUs). This argument holds particularly well whenever precision spatial sampling is not the driver for a given application; using, for example, adaptive binning with SAURON observations \citep{CaCo:03}. However, owing to properties of a datacube, which can be seen as a stack of hundreds of images, IFS with proper spatial sampling has the potential of delivering extremely accurate astrometry with milli-arcsec centroiding precision for point sources from ground-based observations \citep{RoBeKe.:04}. For such applications, where an accurate definition of the point-spread-function is essential, cross-talk corrections are very important (ibid.\ Fig.~4). For the SPIRAL IFU S10 (see Fig.~7 therein) also find unwanted artefacts in their extracted spectra when they do not apply a cross-talk correction.

While spectra of most IFUs are arranged next to each other on both the sky and on the detector, this is not the case for the PPAK IFU. Nearby spectra on the sky are on PPAK placed at different locations on the CCD. When seeing prevents extreme intensity gradients for IFUs with densely packed elements, intensity ratios of as much as 1000 are easily achieved with PPAK data; observing, for example, the center of a planetary nebula and its thousand times fainter halo. In this case it is necessary to use a two-dimensional approach of data reduction in order to extract the weak component properly \citep[cf.][]{BoSc:09}.

\subsection{Comparing the wavelength calibration correction between {\p3d} and {\iraf}}\label{sec:dadm}
Here we compare how well the wavelength calibration correction of {\p3d} matches a correction that is carried out with \textsc{iraf} instead. Using data of the LARR IFU we calculated three dispersion masks, which we then applied to the extracted spectra of the respective arc image. We used the same line list with both tools. With {\p3d} we used a fifth order linear polynomial, and with {\iraf} a fifth order Legendre polynomial. The observations were done using the V300 grating of PMAS, where the dispersion is 1.67\,{\AA}/pixel.

\begin{figure*}
\centering
\includegraphics[width=15cm]{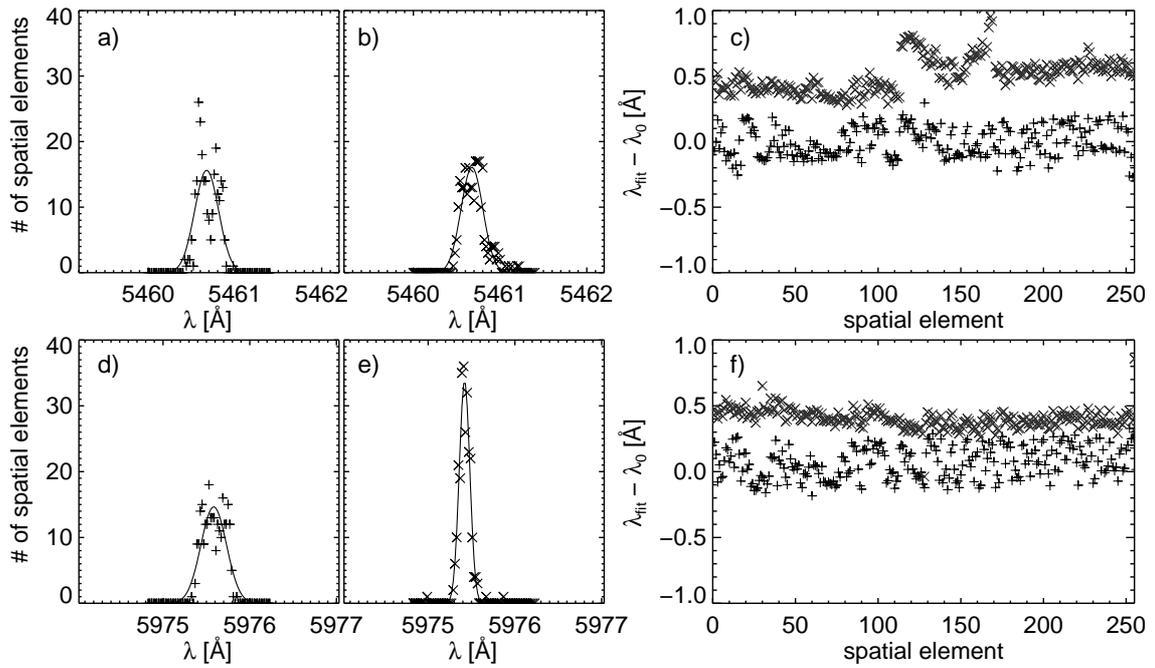}
\caption{This figure shows histograms and residuals of fitted emission lines from two extracted and wavelength-calibrated arc images of LARR data. In the top (bottom) row we show plots of Hg$\,\lambda5461$ (Ne$\,\lambda5976$). Panels \textbf{a} and \textbf{d} (\textbf{b} and \textbf{e}) show histograms of fitted wavelengths using {\p3d} (\iraf), and panels \textbf{c} and \textbf{f} show residuals between the fitted wavelength and $\lambda_0$. The residuals are shown for {\p3d} ($+$) and {\iraf} ($\times$, these residuals are offset by 0.5\,\AA). For further details see Sect.~\ref{sec:dadm}.}
\label{sandinf11}
\end{figure*}

We evaluated the accuracy of either correction by fitting a Gaussian profile to all spectra, using several arc lines in the extracted and wavelength-calibrated images. Specifically we fitted two mercury lines with the rest wavelengths $\lambda_0\!=\!4358.328,\,5460.735\,${\AA}, and two neon lines with $\lambda_0\!=\!5975.534,\,6598.953\,${\AA}. The standard deviation of the center positions were found to be similar for the two tools; they vary between 0.06--0.31\,{\AA} (\iraf) and 0.09--0.13\,{\AA} (\p3d) for the four lines and all spectra of the three images. In Fig.~\ref{sandinf11} we plot histograms of the fitted wavelengths for one of the dispersion masks and Hg$\,\lambda5461$ and Ne$\,\lambda5976$. For the mercury line (Figs.~\ref{sandinf11}a and \ref{sandinf11}b) the respective tool provides the average (and standard deviation) value of 5460.71 (0.15, \iraf) and 5460.69 (0.11, \p3d). The results are evidently very similar in this case. For the neon line the corresponding values are 5975.43 (0.07, \iraf) and 5975.60 (0.12, \p3d). In this case {\p3d} shows a larger scatter of the values. The residual plots (Figs.~\ref{sandinf11}c and \ref{sandinf11}f) confirm that the standard deviation of values of {\iraf} are systematically smaller than with {\p3d}, but differences are small. The wavelength-calibration correction of {\p3d} can likely be further improved if additional care is taken to enhance the resampling algorithm to a common wavelength.

The accuracy that can be achieved with the wavelength calibration depends on several factors. Four such factors are: the spectral resolution, the number of entries in the line mask and the accuracy of the center pixel positions, the fitting function (and its order if it is a polynomial), and properties of the final interpolation to a common wavelength for all spectra. Selecting a line list we have the following four recommendations: lines must be isolated, lines with high S/N are preferred, lines that are (close to) saturated should be avoided, and lines must be distributed across the full spectral range of interest.

\subsection{About the accuracy of spectra which are extracted using the three different methods}\label{sec:daex}
In order to measure the accuracy of the spectrum extraction we test the three extraction methods with idealized simulated data. With this motive we assumed that the line profiles are Gaussian. All spectra are perfectly aligned with the dispersion axis, and the intensity is invariant with wavelength. Hereby we only model one wavelength bin.

Setting up our simulation we first defined a set of ten Gaussian profiles for each of the four IFUs. We used the average instrument-specific profile widths, $G_{\text{mean}}$, which we calculated in Sect.~\ref{sec:dact} (see Table~\ref{sandint3}). The Gaussian profiles are, moreover, separated by $d/\text{bin}_{\dagger}$ pixels (see Table~\ref{sandint2}). We set the intensities of all, but two, profiles to the same value; the third and sixth profile intensities are set ten and hundred times higher. Thereafter the profiles are scaled to a pre-defined value of the signal-to-noise (S/N) of the weaker profiles, and are summed up to create one spectrum. We refer to the scaled intensities as modeled intensities below. For the noise model we used a Poissonian noise distribution and the instrument-specific readout noise. In all subsequent measurements we used the intensity of the second, fifth, and ninth profiles. These profiles correspond to measuring a weaker line that lies next to a line that is 10, 100, and 1 times as strong.

\begin{figure*}
\centering
\includegraphics{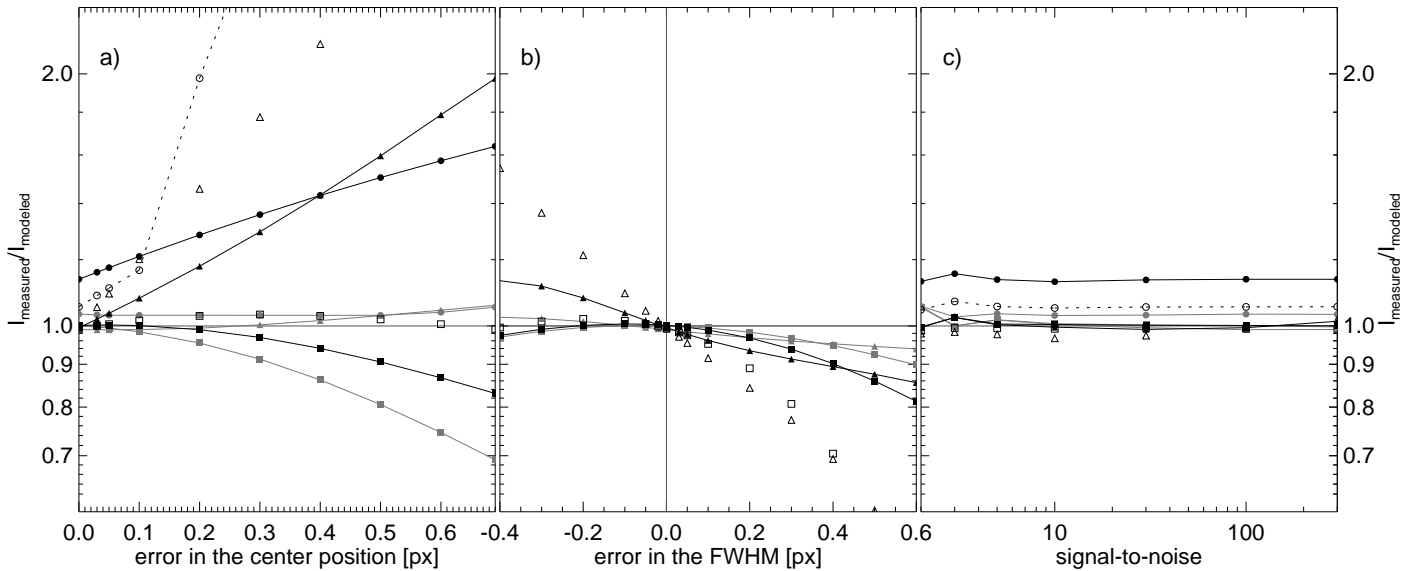}
\caption{This figure shows measured-to-modeled flux ratios for the PPAK IFU (using $\text{bin}_{\dagger}\!=\!2$). The three panels show the flux ratios as a function of: \textbf{a)} the assumed error of the profile center position, \textbf{b)} the assumed error of the profile FWHM and \textbf{c)} the signal-to-noise. We show flux ratios for three sets of profiles where the neighbor profile intensity is stronger by a factor: 1 (gray symbols), 10 (black filled symbols), and 100 (open symbols, for clarity these values are drawn without connecting lines; the exception are the aperture extraction flux ratios which are drawn with a dotted line, these are also offset with -1.0 to keep the values in the same plot). Flux ratios, which are calculated using the standard aperture extraction, are indicated with circles, and ratios of MOX (MPD) with triangles (squares). All values in panels \textbf{a} and \textbf{b} are calculated using S/N=100. The horizontal lines at the flux ratio 1, and the vertical line at error 0.0 in panel \textbf{b}, are guides. For further details see Sect.~\ref{sec:daex}.}
\label{sandinf12}
\end{figure*}

Evaluating the simulation we used an approach that is similar to that of {\rSB}. For each IFU setup we fitted three different sets of Gaussian profiles to the simulated data. For the first set we used the already known center positions and profile widths, and scaled the input intensities to achieve a S/N of 2--300. For the second set we used the known profile width and set the input intensities to correspond to S/N=100, we also introduced an error to the pre-determined profile positions of 0.0--0.7 pixel. In the third set we used the known profile center position and again used S/N=100. This time we introduced an error to the profile width, which are used in the fitting, of -0.4--0.6 pixel; a negative error corresponds to a narrower profile and a positive error to a wider profile. For every configuration we fitted the profiles 100 times using the \textsc{p3d}-routine of the respective extraction method, using as many realizations of the noise model, and saved the average intensities.

In the following text we refer to the modified optimal extraction method as MOX (Sect.~\ref{sec:drspmop}). We, likewise, refer to the multi-profile deconvolution optimal extraction method as MPD (Sect.~\ref{sec:drspmpd}).

We begin our analysis with the PPAK IFU. In Fig.~\ref{sandinf12}a we show the result of the high-intensity simulations where we introduced an error to the pre-determined center positions of the three profiles. Using the aperture extraction it is seen that the extracted flux, due to cross-talk, always is higher than the model input intensity. Note that the values at error 0.0 agree well with the values in Sect.~\ref{sec:dact} (see Table~\ref{sandint3}; we we did not use discrete pixels in that study). The measured flux is $\ga\!100$\% higher than the model flux with the highest intensity ratio. If there are strong intensity gradients in the data, and it is important to measure weak regions accurately, it is not recommended to use the aperture extraction method. Using MOX the accuracy is higher than with the aperture extraction. Although with MOX the error increases rapidly with the error of the pre-determined center position for the two profiles, which lie next to the intenser profiles. MPD shows the best performance of all methods at small errors. Although, for larger errors ($\ga\!0.2\,$px) it is outperformed by the other two methods when the intensity ratio is near unity. With MPD it is enough to keep the profile center error smaller than 0.2\,pixel in order to achieve an intensity error that is smaller than 5\% for all intensity ratios. Using MOX the corresponding error of the pre-determined center position must, in this case, be smaller than about 0.03\,pixel.

We study the influence of an accurate pre-determined profile FWHM in Fig.~\ref{sandinf12}b. Both optimal extraction methods show a similar dependence for positive errors. MPD is seen to give a slightly higher accuracy at reasonably low positive errors ($\la\!0.3\,$pixel). When the pre-determined FWHM is too narrow, and the error consequently is negative, MPD shows smaller errors in the flux than for positive FWHM errors. With MOX the resulting flux errors are similar as for positive FWHM errors. In order to calculate fluxes, which errors are smaller than 5\%, it is necessary to keep the FWHM error smaller than 0.05\,pixel (0.2\,pixel) with MOX (MPD).

The dependence of the resulting flux error on the S/N is shown in Fig.~\ref{sandinf12}c (note that the line of the factor-100 intensity ratio is offset by -1.0 on the ordinate). It is seen that values changes little with the S/N, with the exception of low values (S/N$\la\!4$). Note that for S/N$\la\!30$, and the highest intensity ratio (100), the resulting fluxes, which are calculated using MOX, are about 3\% lower than corresponding values, which are instead calculated using MPD.

\begin{figure*}
\centering
\includegraphics{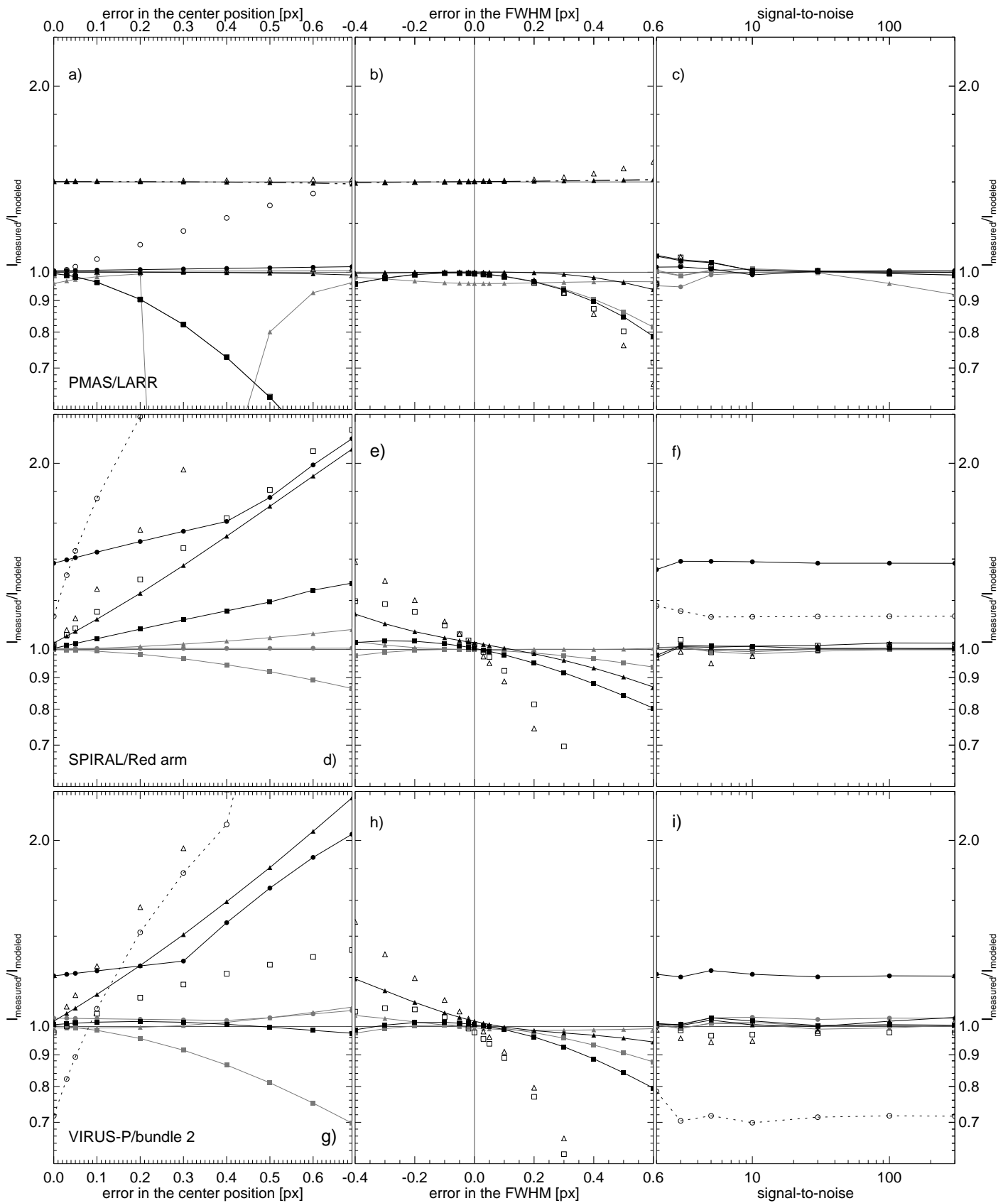}
\caption{Figure~\ref{sandinf12} is repeated for LARR (top row, panels \textbf{a}--\textbf{c}; $\text{bin}_{\dagger}\!=\!2$), SPIRAL (middle row, panels \textbf{d}--\textbf{f}; $\text{bin}_{\dagger}\!=\!1$) and VIRUS-P/bundle~2 (bottom row, panels \textbf{g}--\textbf{i}; $\text{bin}_{\dagger}\!=\!2$). All values in the three panels of the leftmost column are shown as a function of the assumed error of the pre-determined profile center position. Likewise, values in the three panels of the middle (rightmost) column are shown as a function of the assumed error in the pre-determined FWHM (signal-to-noise). In panels \textbf{a} and \textbf{b} we additionally plot flux ratios for MOX (dash-triple-dotted lines which are offset by 0.4\,pixel), for optimal extraction, without cross-talk correction. The dotted lines in panels \textbf{d} and \textbf{f} (\textbf{g} and \textbf{i}) are offset by -3.5 (-2.5).}
\label{sandinf13}
\end{figure*}

In Sect.~\ref{sec:dact} we showed that there is no need to correct LARR data for cross-talk even for high inter-profile intensity ratios, as spectra are very well separated. In order to provide a complete study we, nevertheless, test the outcome when such a correction is made anyways.

Figure~\ref{sandinf13}a shows that MOX gives accurate results for small center position errors and higher intensity gradients. Although when the intensity ratio is unity MOX gives unreliable fluxes, which are $\la\!98\%$ of the true flux for small errors, see Fig.~\ref{sandinf13}b. The offset dash-triple-dotted lines (at flux ratio 1.4) show the calculated fluxes using MOX, without a cross-talk correction -- i.e.\ for regular optimal extraction. The lines illustrate that flux errors are small, even at relatively high errors of the center positions and the FWHM. It does not make any sense to use MPD or MOX with cross-talk correction with LARR data. In this case noise in the profile wings increase the errors of the measured flux, instead of decreasing them.

With the SPIRAL IFU and the VIRUS-P IFU, see Figs.~\ref{sandinf13}d--\ref{sandinf13}i, we see a similar behavior as for the PPAK IFU. With these IFUs and MPD it is necessary to keep errors of the pre-determined center positions $\la\!0.02\,$pixel (SPIRAL) and $\la\!0.15\,$pixel (VIRUS-P) in order to delimit flux errors to 5\%. With MOX the center error should be $\la\!0.01\,$pixel (SPIRAL) and $<\!0.03\,$pixel (VIRUS-P). The required precision of center positions is very high for SPIRAL, with both methods, which is why it seems unrealistic to achieve this high accuracy in fluxes. In order to delimit errors we recommend to fit the spectrum center positions, along with the intensities, when calculating line profiles for this IFU (cf.\ Sect.~\ref{sec:datm}). Both methods, moreover, require an accuracy in the profile width of $\la\!0.05\,$pixel in order to constrain the measured flux error to 5\%. The S/N-test for VIRUS-P (Fig.~\ref{sandinf13}i) shows that the error of MPD is about half that of MOX for S/N$\la\!10$. For the SPIRAL IFU significant flux errors ($\la\!4\%$) are only introduced, as a function of S/N, with MOX (Fig.~\ref{sandinf13}f). 

We conclude that MPD is the preferred method of spectrum extraction whenever cross-talk is present. The cases where MOX was found to outperform MPD can likely be explained by the smaller extraction width that is used with MOX (compare $\overline{x}_{\text{w}}$ with $\hat{x}_{\text{w}}$ that were both defined in Sect.~\ref{sec:drspextract}). Regardless of the chosen method of optimal extraction it is always important to minimize errors of the pre-determined center positions and FWHM. Errors of measured fluxes grow fast if there are significant intensity gradients in the data. We found that MOX is highly sensitive to accurate spectrum center positions (i.e.\ traces). MOX also introduces some error to data with low S/N. Although with LARR data, which is free of cross-talk, MOX is the preferred method (because of the way the extraction width of both methods is defined within \p3d), without the optional cross-talk correction. Finally, with the exception of LARR data, aperture extraction always introduces significant errors to calculated fluxes of data, which contains any spatial intensity gradients.

\begin{figure*}
\centering
\includegraphics{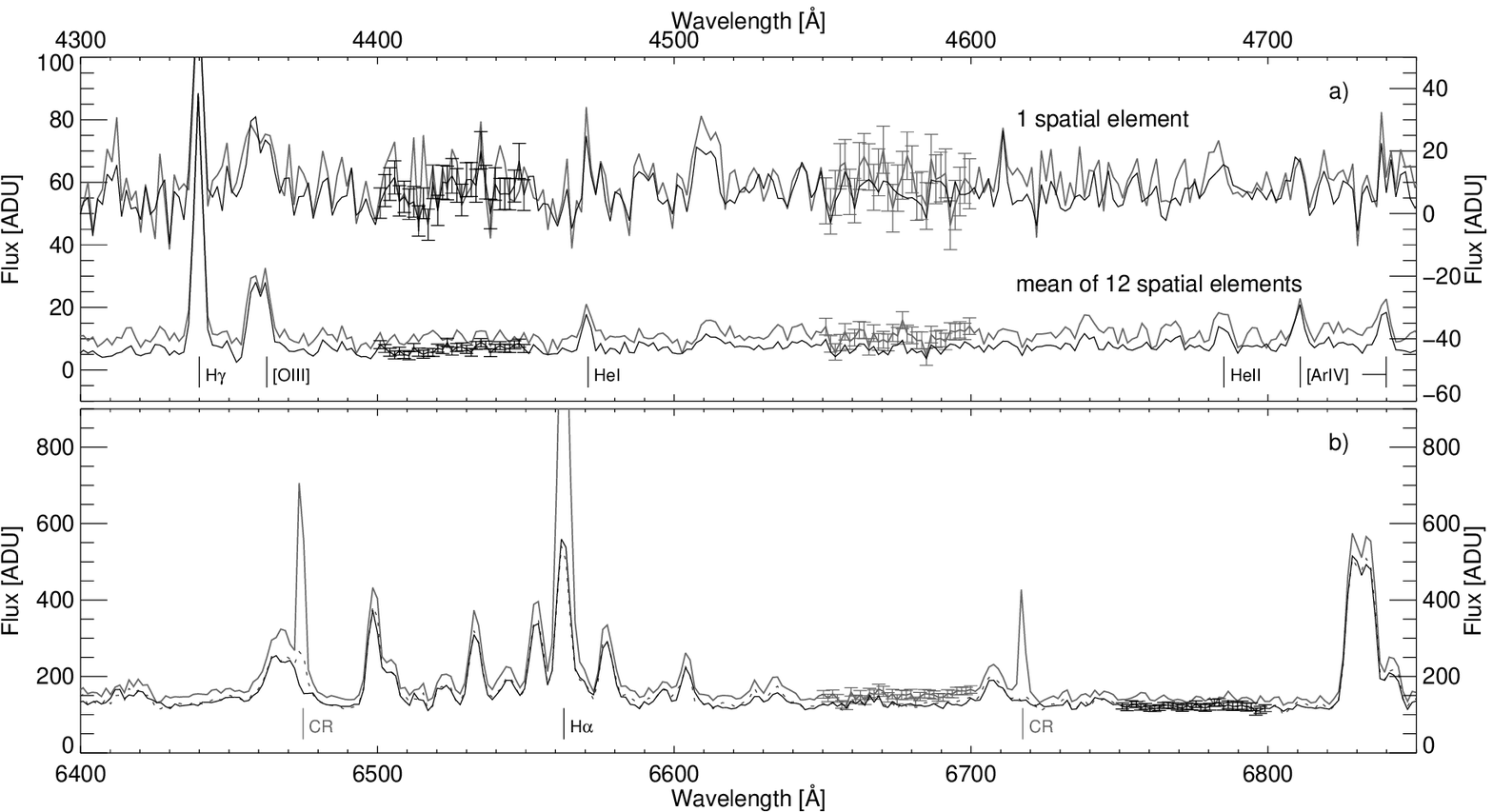}
\caption{Comparison of spectra, which are calculated using the aperture extraction, MOX and MPD. In panel \textbf{a} we show spectra of LARR and in panel \textbf{b} spectra of PPAK. The flux is shown as a function of wavelength -- note that the two panels are plotted for different wavelength ranges. Solid black (gray) lines show spectra using MOX (aperture extraction). In panel \textbf{b} we used MPD to calculate the spectrum that is drawn with a dotted line. In panel \textbf{a} the two lower (upper) spectra use the left (right) y-axis. Error bars are shown for parts of the spectra in both panels. Wavelengths of a few emission lines (cosmic ray hits) are indicated with a vertical line and the name of the line (CR). For further details see Sect.~\ref{sec:daoo}.}
\label{sandinf14}
\end{figure*}

\begin{figure}
\centering
\includegraphics{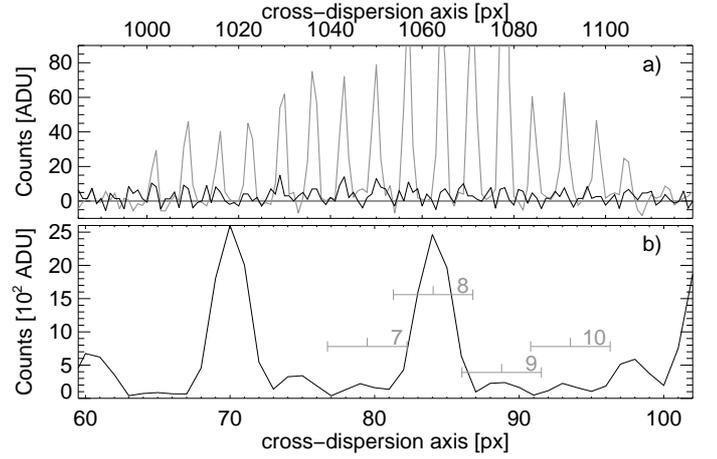}
\caption{We plot parts of the bias-subtracted raw data, which we used to extract the spectra of Fig.~\ref{sandinf14}. The counts are plotted for one wavelength across the cross-dispersion axis for data of LARR, panel \textbf{a}, and PPAK, panel \textbf{b}. In panel \textbf{a} the group of profiles drawn in gray is shown for the center wavelength of H$\gamma$; the black line is drawn for the corresponding wavelength bin 5 pixels towards the blue of the center of H$\gamma$. The dark gray horizontal line marks the zero-level. In panel \textbf{b} similar profiles are shown about the center wavelength of H$\alpha$. Positions of four apertures are shown with gray indicators. For further details see Sect.~\ref{sec:daoo}.}
\label{sandinf15}
\end{figure}

\subsection{Comparing resulting emission line spectra of planetary nebulae using the three different extraction methods}\label{sec:daoo}
Finally, we compare the outcome of spectra, which were extracted using the different methods. In Fig.~\ref{sandinf14} we show wavelength-calibrated and flat-fielded spectra of data of LARR and PPAK from the central regions of two planetary nebulae (NGC\,2392 for PPAK, and M\,2-2 for LARR, which outcome is presented in \citealt{SaScRo.:08}). In the first case we reduced the LARR data using MOX (without correcting for cross-talk) and aperture extraction. As Fig.~\ref{sandinf14}a shows the spectrum of a single spatial element is less noisy when MOX is used instead of aperture extraction. The error bars of MOX are half as high than with aperture extraction. In the spectrum that is averaged over 12 spatial elements we see that the continuum level of the MOX-spectrum is lower than that of the aperture-extracted spectrum (by about 55\%). In this emission line object scattered light becomes significant for the weak continuum that is dominated by readout noise (compare the gray and black lines in Fig.~\ref{sandinf15}a). Using aperture extraction more of the scattered light is included in the resulting spectrum, and weak emission lines are underestimated. Note that both methods calculate the same intensity for H$\gamma$.

In order to illustrate the result of cross-talk in PPAK data we show Fig.~\ref{sandinf14}b. In this case MOX and MPD give very similar solutions, except at the location of the cosmic ray hit, at $\lambda\!\simeq\!6474\,${\AA}. The flux at the wavelength of the cosmic ray hit was removed by the MOX-algorithm. The shown spectrum lies in a weak part of the central nebula on the sky, but on the CCD the neighbor spectrum lies in a more intense part of the nebula; compare the apertures 9 and 8 in Fig.~\ref{sandinf15}b. Using aperture extraction flux of the neighbor spectrum falls within the aperture of the weak spectrum, and increases the flux of H$\alpha$ significantly above its true value. In this case we attribute the difference in the continuum levels of the aperture and MOX/MPD methods to both scattered light and cross-talk. The continuum-region error bars of MOX are on average 0.88 times as high as the aperture extraction error bars.

\section{Conclusions}\label{sec:conclusions}
We have presented a new and general data-reduction tool for fiber-fed IFUs. Our goal has been to write a user-friendly tool that works stably with different sets of instruments and data. In comparison to similar tools a strong advantage of {\p3d} is that it can find and trace all spectra on the detector, mostly without any user interaction at all. Using the methods of optimal spectrum extraction data of all IFUs can also be corrected for cross-talk, which arises due to overlapping spectra on the detector. Since the same procedures are used with all implemented IFUs it is, moreover, a straightforward task to compare the outcome of different observations. Although {\p3d} is based on the proprietary Interactive Data Language (IDL) its use requires no IDL license. All components of {\p3d} work with all platforms supported by IDL. The program code can be downloaded from the project web site at \anchor{http://p3d.sourceforge.net}{http://p3d.sourceforge.net}.

In order to validate the program code we have tested the different parts using both simulated data and corresponding outcome of \iraf. We found that {\p3d} produces results comparable to {\iraf}. For all IFUs, with the exception of the lens array of PMAS, {\p3d} is able to extract more accurate values than {\iraf} since {\p3d} can correct for cross-talk.

Although {\p3d} has so far been configured for four IFUs it is a straightforward task to extend it to work with additional instruments. If the new IFU is similar to the already implemented ones it is just a matter of setting up another set of instrument-specific parameters. The same concerns the level of functionality. {\p3d} can with relatively small effort be extended to also handle, for example, flux calibration, correction for differential atmospheric refraction (for those IFUs where it makes sense), removal of cosmic rays in individual images, sky subtraction, and account for scattered light. Most parts of {\p3d} are, moreover, fast and only require on the order of seconds to execute on a typical workstation. The calculation of line profiles and the optimal extraction algorithm, however, are more computationally intensive and require on the order of a few minutes. The code execution time could in this case be shortened by moving the relevant parts of the IDL-code to compiled (and dynamically loaded) C-code. If, and when, these and other improvements will be implemented depends on the need and the interest of the community.

\begin{acknowledgements}
During a large part of the development time this project was supported by the grants BMBF\,03Z2A51 and BMBF\,05A08BA1. A.~M.-I.\ was supported by the Spanish Ministry of Science and Innovation (MICINN) under the program ``Specialization in International Organisms'' (ES2006-0003). We thank L.~Cair{\'o}s for her careful testing of the dispersion mask tool, and for providing us with test data sets for bundle~1 of VIRUS-P. We thank O.~Streicher both for his careful testing of all parts of {\p3d}, and for having written tools for the project web site, which automatically extracts and presents the routine documentation. Q.~Parker and M.~Rela\~no are thanked for providing us with test data sets for the LARR and SPIRAL IFUs. We, furthermore, thank A.~Zwanzig for writing extensive installation instructions. We finally thank J.~Adams and R.~Sharp for providing us with instrument-specific information, which allowed us to configure {\p3d} for VIRUS-P and SPIRAL, respectively.
\end{acknowledgements}


\begin{thebibliography}{29}
\expandafter\ifx\csname natexlab\endcsname\relax\def\natexlab#1{#1}\fi

\bibitem[{{Adams} {et~al.}(2010)}]{Adetal:10}
{Adams}, J., {et~al.} 2010, in prep.\

\bibitem[{{Allington-Smith} \& {Content}(1998)}]{AlCo:98}
{Allington-Smith}, J. \& {Content}, R. 1998, \pasp, 110, 1216

\bibitem[{{Allington-Smith} {et~al.}(2002){Allington-Smith}, {Murray},
  {Content}, {Dodsworth}, {Davies}, {Miller}, {Jorgensen}, {Hook}, {Crampton},
  \& {Murowinski}}]{AlMuCo.:02}
{Allington-Smith}, J., {Murray}, G., {Content}, R., {et~al.} 2002, \pasp, 114,
  892

\bibitem[{{Alonso-Herrero} {et~al.}(2009){Alonso-Herrero},
  {Garc{\'{\i}}a-Mar{\'{\i}}n}, {Monreal-Ibero}, {Colina}, {Arribas},
  {Alfonso-Garz{\'o}n}, \& {Labiano}}]{AlGaMo.:09}
{Alonso-Herrero}, A., {Garc{\'{\i}}a-Mar{\'{\i}}n}, M., {Monreal-Ibero}, A.,
  {et~al.} 2009, \aap, 506, 1541

\bibitem[{{Arribas} {et~al.}(1998){Arribas}, {Carter}, {Cavaller}, {del Burgo},
  {Edwards}, {Fuentes}, {Garcia}, {Herreros}, {Jones}, {Mediavilla}, {Pi},
  {Pollacco}, {Rasilla}, {Rees}, \& {Sosa}}]{ArCaCa.:98}
{Arribas}, S., {Carter}, D., {Cavaller}, L., {et~al.} 1998, in Proc.\ SPIE, ed.
  S.~{D'Odorico}, Vol. 3355, 821

\bibitem[{{Avila} {et~al.}(2003){Avila}, {Guinouard}, {Jocou}, {Guillon}, \&
  {Balsamo}}]{AvGuJo.:03}
{Avila}, G., {Guinouard}, I., {Jocou}, L., {Guillon}, F., \& {Balsamo}, F.
  2003, in Proc.\ SPIE, ed. {M.~Iye \& A.~F.~M.~Moorwood}, Vol. 4841, 997--1005

\bibitem[{{Becker}(2002)}]{TBe:02}
{Becker}, T. 2002, PhD thesis, Univ.\ Potsdam

\bibitem[{{Bershady}(2009)}]{Be:09}
{Bershady}, M.~A. 2009, in 3D Spectroscopy in Astronomy, XVII Canary Island
  Winter School of Astrophysics, ed. E.~{Mediavilla}, S.~{Arribas}, M.~{Roth},
  J.~{Cepa-Nogue}, \& F.~{S{\'a}nchez} ({CUP})

\bibitem[{{Blecha} {et~al.}(2000){Blecha}, {Cayatte}, {North}, {Royer}, \&
  {Simond}}]{BlCaNo.:00}
{Blecha}, A., {Cayatte}, V., {North}, P., {Royer}, F., \& {Simond}, G. 2000, in
  Proc.\ SPIE, ed. M.~{Iye} \& A.~F. {Moorwood}, Vol. 4008, 467

\bibitem[{{Bolton} \& {Burles}(2007)}]{BoBu:07}
{Bolton}, A.~S. \& {Burles}, S. 2007, New Journal of Physics, 9, 443

\bibitem[{{Bolton} \& {Schlegel}(2009)}]{BoSc:09}
{Bolton}, A.~S. \& {Schlegel}, D.~J. 2009, \pasp, submitted (arxiv:0911.2689)

\bibitem[{{Cappellari} \& {Copin}(2003)}]{CaCo:03}
{Cappellari}, M. \& {Copin}, Y. 2003, \mnras, 342, 345

\bibitem[{{Fabrika} {et~al.}(2005){Fabrika}, {Sholukhova}, {Becker},
  {Afanasiev}, {Roth}, \& {S{\'a}nchez}}]{FaShBe.:05}
{Fabrika}, S., {Sholukhova}, O., {Becker}, T., {et~al.} 2005, \aap, 437, 217

\bibitem[{{Filippenko}(1982)}]{Fi:82}
{Filippenko}, A.~V. 1982, \pasp, 94, 715

\bibitem[{{Hill} {et~al.}(2008){Hill}, {MacQueen}, {Smith}, {Tufts}, {Roth},
  {Kelz}, {Adams}, {Drory}, {Grupp}, {Barnes}, {Blanc}, {Murphy}, {Altmann},
  {Wesley}, {Segura}, {Good}, {Booth}, {Bauer}, {Popow}, {Goertz}, {Edmonston},
  \& {Wilkinson}}]{HiMaSm.:08}
{Hill}, G.~J., {MacQueen}, P.~J., {Smith}, M.~P., {et~al.} 2008, in Proc.\
  SPIE, Vol. 7014, 231

\bibitem[{{Horne}(1986)}]{Ho:86}
{Horne}, K. 1986, \pasp, 98, 609 (\rH)

\bibitem[{{Howell}(2006)}]{Ho:06}
{Howell}, S.~B. 2006, {Handbook of CCD Astronomy}, 2nd edn., {Cambridge
  Observing Handbooks for Research Astronomers} ({CUP})

\bibitem[{{Hunt} \& {Thomas}(1999)}]{HuTh:99}
{Hunt}, A. \& {Thomas}, D. 1999, {The Pragmatic Programmer: From Journeyman to
  Master} (Addison-Wesley Professional)

\bibitem[{{Kelz} {et~al.}(2006){Kelz}, {Verheijen}, {Roth}, {Bauer}, {Becker},
  {Paschke}, {Popow}, {S{\'a}nchez}, \& {Laux}}]{KeVeRo.:06}
{Kelz}, A., {Verheijen}, M.~A.~W., {Roth}, M.~M., {et~al.} 2006, \pasp, 118,
  129

\bibitem[{{Kissler-Patig} {et~al.}(2004){Kissler-Patig}, {Copin}, {Ferruit},
  {P{\'e}contal-Rousset}, \& {Roth}}]{KiCoFe.:04}
{Kissler-Patig}, M., {Copin}, Y., {Ferruit}, P., {P{\'e}contal-Rousset}, A., \&
  {Roth}, M.~M. 2004, Astronomische Nachrichten, 325, 159

\bibitem[{{LeF{\`e}vre} {et~al.}(2003){LeF{\`e}vre}, {Saisse}, {Mancini}, {Brau-Nogue},
  {Caputi}, {Castinel}, {D'Odorico}, {Garilli}, {Kissler-Patig}, {Lucuix},
  {Mancini}, {Pauget}, {Sciarretta}, {Scodeggio}, {Tresse}, \&
  {Vettolani}}]{LeSaMa.:03}
{LeF{\`e}vre}, O., {Saisse}, M., {Mancini}, D., {et~al.} 2003, in Proc.\ SPIE, ed.
  M.~{Iye} \& A.~F.~M. {Moorwood}, Vol. 4841, 1670

\bibitem[{{Lehmann} {et~al.}(2005){Lehmann}, {Becker}, {Fabrika}, {Roth},
  {Miyaji}, {Afanasiev}, {Sholukhova}, {S{\'a}nchez}, {Greiner}, {Hasinger},
  {Costantini}, {Surkov}, \& {Burenkov}}]{LeBeFa.:05}
{Lehmann}, I., {Becker}, T., {Fabrika}, S., {et~al.} 2005, \aap, 431, 847

\bibitem[{{Markwardt}(2009)}]{Ma:09}
{Markwardt}, C.~B. 2009, in ASP Conf.\ Series, Vol. 411, Astronomical Data
  Analysis Software and Systems XVIII, ed. D.~{Bohlender}, P.~{Dowler}, \&
  D.~{Durand}, 251

\bibitem[{{Monreal-Ibero} {et~al.}(2005){Monreal-Ibero}, {Roth},
  {Sch{\"o}nberner}, {Steffen}, \& {B{\"o}hm}}]{MoRoSc.:05}
{Monreal-Ibero}, A., {Roth}, M.~M., {Sch{\"o}nberner}, D., {Steffen}, M., \&
  {B{\"o}hm}, P. 2005, \apjl, 628, L139

\bibitem[{{Pasquini} {et~al.}(2000){Pasquini}, {Avila}, {Allaert}, {Ballester},
  {Biereichel}, {Buzzoni}, {Cavadore}, {Dekker}, {Delabre}, {Ferraro}, {Hill},
  {Kaufer}, {Kotzlowski}, {Lizon}, {Longinotti}, {Moureau}, {Palsa}, \&
  {Zaggia}}]{PaAvAl.:00}
{Pasquini}, L., {Avila}, G., {Allaert}, E., {et~al.} 2000, in Proc.\ SPIE, ed.
  M.~{Iye} \& A.~F. {Moorwood}, Vol. 4008, 129

\bibitem[{{Pych}(2004)}]{Py:04}
{Pych}, W. 2004, \pasp, 116, 148

\bibitem[{{Rela\~no} {et~al.}(2009){Rela\~no}, {Monreal-Ibero}, {V{\'{\i}}lchez}, \&
  {Kennicutt}}]{ReMoVi.:09}
{Rela\~no}, M., {Monreal-Ibero}, A., {V{\'{\i}}lchez}, J.~M., \& {Kennicutt}, R.~C.
  2009, MNRAS, 402, 1635

\bibitem[{{Roth} {et~al.}(2004){Roth}, {Becker}, {Kelz}, \&
  {Schmoll}}]{RoBeKe.:04}
{Roth}, M.~M., {Becker}, T., {Kelz}, A., \& {Schmoll}, J. 2004, \apj, 603, 531

\bibitem[{{Roth} {et~al.}(2005){Roth}, {Kelz}, {Fechner}, {Hahn}, {Bauer},
  {Becker}, {B{\"o}hm}, {Christensen}, {Dionies}, {Paschke}, {Popow}, {Wolter},
  {Schmoll}, {Laux}, \& {Altmann}}]{RoKeFe.:05}
{Roth}, M.~M., {Kelz}, A., {Fechner}, T., {et~al.} 2005, \pasp, 117, 620

\bibitem[{{S{\'a}nchez}(2006)}]{Sa:06}
{S{\'a}nchez}, S.~F. 2006, Astronomische Nachrichten, 327, 850 (\rS)

\bibitem[{{Sandin} {et~al.}(2008){Sandin}, {Sch\"onberner}, {Roth}, {Steffen},
  {B\"ohm}, \& {Monreal-Ibero}}]{SaScRo.:08}
{Sandin}, C., {Sch\"onberner}, D., {Roth}, M.~M., {et~al.} 2008, \aap, 486, 545

\bibitem[{{Schmoll} {et~al.}(2004){Schmoll}, {Dodsworth}, {Content}, \&
  {Allington-Smith}}]{ScDoCo.:04}
{Schmoll}, J., {Dodsworth}, G.~N., {Content}, R., \& {Allington-Smith}, J.~R.
  2004, in Proc.\ SPIE, ed. A.~F.~M. {Moorwood} \& M.~{Iye}, Vol. 5492,
  624

\bibitem[{{Scodeggio} {et~al.}(2005){Scodeggio}, {Franzetti}, {Garilli},
  {Zanichelli}, {Paltani}, {Maccagni}, {Bottini}, {Le Brun}, {Contini},
  {Scaramella}, {Adami}, {Bardelli}, {Zucca}, {Tresse}, {Ilbert}, {Foucaud},
  {Iovino}, {Merighi}, {Zamorani}, {Gavignaud}, {Rizzo}, {McCracken}, {Le
  F{\`e}vre}, {Picat}, {Vettolani}, {Arnaboldi}, {Arnouts}, {Bolzonella},
  {Cappi}, {Charlot}, {Ciliegi}, {Guzzo}, {Marano}, {Marinoni}, {Mathez},
  {Mazure}, {Meneux}, {Pell{\`o}}, {Pollo}, {Pozzetti}, \&
  {Radovich}}]{ScFrGa.:05}
{Scodeggio}, M., {Franzetti}, P., {Garilli}, B., {et~al.} 2005, \pasp, 117,
  1284

\bibitem[{{Sharp} \& {Birchall}(2010)}]{ShBi:10}
{Sharp}, R. \& {Birchall}, M.~N. 2010, \pasa, in press (arxiv:0912.0558, \rSB)

\bibitem[{{Sharp} {et~al.}(2006){Sharp}, {Saunders}, {Smith}, {Churilov},
  {Correll}, {Dawson}, {Farrel}, {Frost}, {Haynes}, {Heald}, {Lankshear},
  {Mayfield}, {Waller}, \& {Whittard}}]{ShSaSm.:06}
{Sharp}, R., {Saunders}, W., {Smith}, G., {et~al.} 2006, in Proc.\ SPIE, Vol.
  6269

\bibitem[{{Smith} {et~al.}(2004){Smith}, {Saunders}, {Bridges}, {Churilov},
  {Lankshear}, {Dawson}, {Correll}, {Waller}, {Haynes}, \&
  {Frost}}]{SmSaBr.:04}
{Smith}, G.~A., {Saunders}, W., {Bridges}, T., {et~al.} 2004, in Proc.\ SPIE,
  ed. A.~F.~M. {Moorwood} \& M.~{Iye}, Vol. 5492, 410

\bibitem[{{Turner}(2006)}]{Tu:06}
{Turner}, J.~E.~H. 2006, New Astronomy Review, 50, 392

\bibitem[{{van Dokkum}(2001)}]{vDo:01}
{van Dokkum}, P.~G. 2001, \pasp, 113, 1420

\bibitem[{{Villar-Mart{\'{\i}}n} {et~al.}(2006){Villar-Mart{\'{\i}}n},
  {S{\'a}nchez}, {De Breuck}, {Peletier}, {Vernet}, {Rettura}, {Seymour},
  {Humphrey}, {Stern}, {di Serego Alighieri}, \& {Fosbury}}]{ViSaDe.:06}
{Villar-Mart{\'{\i}}n}, M., {S{\'a}nchez}, S.~F., {De Breuck}, C., {et~al.}
  2006, \mnras, 366, L1

\end{thebibliography}
\end{document}